\documentclass[%
reprint,
superscriptaddress,
 amsmath,amssymb,
prb,
]{revtex4-2}

\usepackage{graphicx}
\graphicspath{ {./figures/} }
\usepackage{dcolumn}
\usepackage{bm}
\usepackage[colorlinks, allcolors=blue]{hyperref}

\usepackage[english]{babel}
\usepackage{amssymb}
\usepackage{lipsum}
\usepackage[utf8]{luainputenc}
\usepackage{indentfirst}
\usepackage{microtype}
\usepackage[version=4]{mhchem}
\usepackage{physics}
\usepackage{amsmath,amssymb}
\usepackage{bm}
\usepackage{float}
\usepackage{floatflt}
\usepackage{bbold}
\usepackage{chemformula}

\begin{document}


\title{Two-phonon pairing and superconductivity in \ch{SrTiO3}}


\author{Antonio Santacesaria} 
\affiliation{Dipartimento di Fisica, Sapienza Università di Roma, 00185 Rome, Italy}
\affiliation{ISC-CNR, Istituto dei Sistemi Complessi, via dei Taurini 19, 00185 Rome, Italy}%

\author{Cristiano Muzzi}
\affiliation{International Solvay Institutes, 1050 Brussels, Belgium}
\affiliation{Center for Nonlinear Phenomena and Complex Systems, Universit\'e Libre de Bruxelles, CP 231, Campus Plaine, B-1050 Brussels, Belgium}

\author{Maria Eleonora Temperini}%
\affiliation{Dipartimento di Fisica, Sapienza Università di Roma, 00185 Rome, Italy}

\author{Paolo Barone}
\affiliation{SPIN-CNR, Istituto Superconduttori, Materiali Innovativi e Dispositivi, Area della Ricerca di Tor Vergata, via del Fosso del Cavaliere 100, 00133 Rome, Italy}	

\author{Maria N. Gastiasoro} 
\affiliation{Donostia International Physics Center, Donostia-San Sebastian, Manuel Lardizabal Ibilbidea, 4, 20018 Spain}%

  \author{Jos\'{e} Lorenzana}
 \email{jose.lorenzana@cnr.it}
\affiliation{ISC-CNR, Istituto dei Sistemi Complessi, via dei Taurini 19, 00185 Rome, Italy}%
\affiliation{Dipartimento di Fisica, Sapienza Università di Roma, 00185 Rome, Italy}

\date{\today}

\begin{abstract}
We explore the possibility that the two-phonon pairing mechanism proposed by Ngai can explain superconductivity in doped strontium titanate (STO).
The two-phonon deformation potential is evaluated using two distinct theoretical estimates based on first-principles calculations and two empirical estimates based on experimental data, yielding very large values of the same order of magnitude. We derive the effective electron-electron interaction mediated by two-phonon exchange. Crucial to our computations is the strong dispersion of the involved soft phonons. 
Because the scale of the two-phonon interaction is much larger than the Fermi energy, the Migdal theorem does not apply. 
We obtain a one-loop Eliashberg equation that can be solved in the weak-coupling limit. We find that despite the significant electron-phonon matrix element, phase space considerations considerably reduce the effective coupling. A two-phonon mechanism can not be excluded based on the $T_c$ magnitude, but its value is dominated by the high-energy-frequency physics and is weakly affected by the soft-phonon behavior. The theory predicts a $T_c$ that monotonously increases with doping, which is not in accordance with the experiment. 
We identify the conditions for a two-phonon mechanism to be effective in other materials and discuss possible candidates. 
\end{abstract}

\maketitle





\section{Introduction}
\label{introduction}
Strontium titanate (STO) is a band insulator whose dielectric constant tends to diverge at low temperatures.  Typically, this behavior signals a ferroelectric (FE) instability which, however, in STO is believed to be avoided due to quantum fluctuations~\cite{Epsilon2,Littlewood2023}. The proximity to the instability manifests also through the softening of two
transverse optical (TO) phonons. Small perturbations such as isotope substitution, strain, or doping can drive the transition to a long-range ordered ferroelectric phase~\cite{Collignon2019,Gastiasoro2020Review}, therefore, STO can be said to be very close to a ferroelectric quantum critical point.  

The doped phase of STO is also very peculiar as it is one of the lowest density superconductors existing in nature, with a charge carrier density in the range $10^{17}-10^{21}\text{cm}^{-3}$. Since the
Debye frequency is larger than the Fermi energy, in principle the Migdal theorem does not apply, which makes the use of BCS theory contentious. A key question in this context is which boson is the mediator of the pairing interaction. 
Both theory~\cite{Edge2015,Wolfe2018} and  experiments~\cite{Stucky2016N,Enderlein2020,Ahadi2019N,Russell2019N,Franklin2021,Rischau2022} 
point to the role of the proximity to a ferroelectric instability in the superconducting phenomena.
Of course, once the system is metallic, it can not be ferroelectric in the literal sense, but we follow the common practice to call the metallic phase with broken inversion symmetry a ``ferroelectric metal" \cite{Anderson_PRL_ferroelectricmetal1965,Klein2023}. 

In the conventional electron-phonon mechanism, electrons couple to longitudinal phonons~\cite{Mahan2000} and not to transverse modes. Crucially,  longitudinal optical modes (LO) in STO are protected from softening by the long-range Coulomb interaction, which is weakly affected by free-carrier screening given the low charge-carrier densities relevant for doped STO, while only TO modes become soft at a ferroelectric instability. 
Therefore, there has been a search for alternative electron-phonon coupling mechanisms with transverse modes that can provide pairing and hopefully explain superconductivity. 
\begin{figure}[tb]
\centering  
\includegraphics[width=1.1\linewidth]{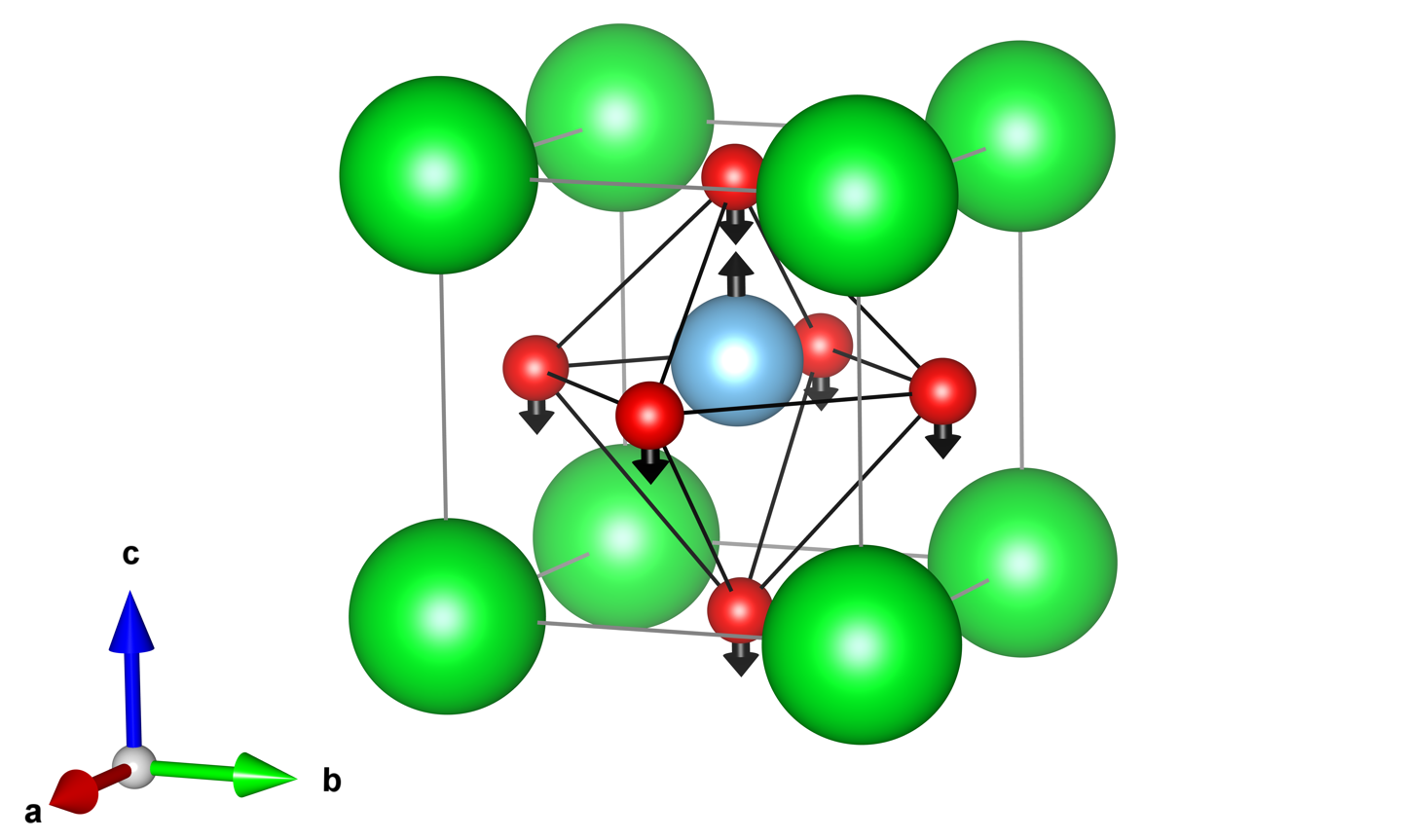}
    \caption{\ch{SrTiO3} cubic unit cell. We show oxygen atoms in red, titanium atoms in blue, and strontium atoms in green. Black arrows represent the Slater mode displacements, for which oxygen atoms move coherently as a cage with opposite phases with respect to the titanium atom while strontium atoms are fixed in their equilibrium position. Notice that DFT calculations have been made with the tetragonal unit cell.}
    \label{fig:Slater}
\end{figure}

Two main mechanisms have been invoked for STO and the related material \ch{KTaO3} (KTO): a one-phonon dynamical Rashba-like mechanism assisted by spin-orbit coupling~\cite{Gastiasoro2020Anisotropic,Gastiasoro2022,Gastiasoro2023,NOrman2023,Venditti2023,Venditti2025,Norman2026,Klein2023} and a two-phonon mechanism first proposed by Ngai~\cite{Ngai,VanderMarel2019,Kiselov2021N,Volkov2022,Saha2025}.
Previously, some of us have used density functional theory (DFT) to estimate the one-phonon matrix element and explored the Rashba mechanism in STO~\cite{Gastiasoro2022,Gastiasoro2023} and KTO~\cite{Venditti2023,Venditti2025}. Here, we use the same DFT computations to estimate the electron-phonon matrix element for the Ngai~\cite{Ngai} mechanism involving two transverse soft phonons assumed to be of the Slater form (Fig.~\ref{fig:Slater}). Furthermore, 
we derive the electron-two-phonon interaction Hamiltonian and the effective two-phonon-mediated electron-electron coupling for STO. A crucial role in our computations is played by the strong dispersion of the TO modes, which are soft only close to the $\Gamma$ point of the Brillouin zone. This is in contrast to some recent works~\cite{Han2024N,Zappacosta2025,Ragni2023} which considered the quadratic electron-phonon coupling with a dispersionless phonon mode. 

Our frozen-phonon computations reveal that for a displacement of the order of the zero-point motion~\cite{Gastiasoro2022} of the soft mode ($l_s\approx 0.3$~\AA) or even much smaller, the bands are much more affected by the term quadratic in the displacement than the Rashba-like linear one (Fig.~\ref{fig:DFT}). This suggests a gigantic electron-two-phonon interaction, which may lead to superconductivity.  Unfortunately, a careful analysis within a one-loop Eliashberg theory shows that while a contribution to superconductivity can not be excluded, it does not appear to be dominant.  We discuss the conditions for two-phonon superconductivity in other materials.

\section{Model and Methods}\label{an}

The high-temperature structure of STO is shown in Fig.~\ref{fig:Slater}. Each $\ce{Ti}$ atom is situated at the center of
a $\ce{TiO_6}$ octahedron and each $\ce{Sr}$ atom is situated at the center of an $\ce{SrO_{12}}$ cuboctahedron. 
Below 105 K, STO has an antiferrodistortive (AFD) tetragonal structure with 
space group I4/mcm. For simplicity, we will index directions using pseudocubic coordinates. Figure \ref{fig:tetra} in Appendix~\ref{app} shows the relation between the tetragonal unit cell and the pseudocubic unit cell. 
Figures~\ref{fig:Slater}, \ref{fig:tetra} also show the displacement pattern of the Slater mode for the TO phonon, which will be discussed below.

We performed DFT computations using the projector augmented-wave (PAW) method as implemented in VASP~\cite{Kresse1996,Kresse1999}, employing the Perdew-Burke-Ernzerhof~\cite{Perdew2008} generalized gradient approximation revised for solids (PBEsol).
We initially relaxed the tetragonal structures until the forces were below 1 meV/Å, utilizing a plane-wave cutoff of 520 eV and a Monkhorst-Pack grid of $6 \times 6 \times 6$ $k$-points. The optimized tetragonal unit cell lattice constants were determined as  $a^T =b^T= 5.508$ Å and $c^T = 7.845$ Å, corresponding to a (pseudo)cubic lattice constant $a=3.895$ Å and a small tetragonality factor $c/a-1=0.007$. Subsequently, electronic structure calculations were performed with the inclusion of spin-orbit coupling (SOC) in VASP, which is known to have a strong impact on energy splittings of the order or larger than typical Fermi energies. 

The low-energy electronic band structure of the equilibrium phase is depicted as black lines in Fig.~\ref{fig:DFT}. Because of inversion symmetry, it consists of three spin-degenerate bands around the zone center that can be ascribed to Ti $3d$ $t_{2g}$ orbitals in the cubic phase, where they would be fully degenerate at $\Gamma$ in the absence of relativistic corrections. As reported, e.g., in Ref. \cite{Gastiasoro2023}, the band splitting at $\Gamma$ originates from the interplay of SOC and tetragonal AFD distortion. We label the bands with integer numbers starting from the lower band (1).

\subsection{Model}

\begin{figure}[tb]
    \includegraphics[width=0.7\linewidth]{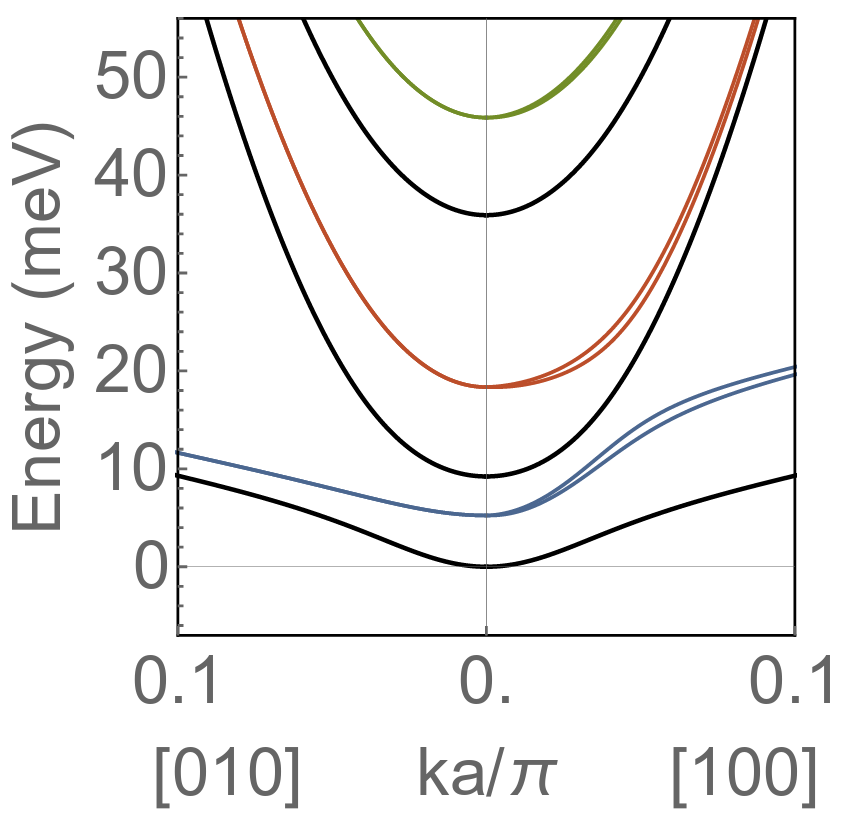}
    \caption{Equilibrium doubly degenerate bands (black) and band structure in the presence of a Slater frozen phonon ($1$ blue, $2$ red and $3$ green).
    Directions are indexed in pseudocubic coordinates. 
    The lattice displacement is along [010] with magnitude ${ u}_S = 0.02 ~\text{\AA}$. 
    Notice that this displacement is much smaller than the estimated zero-point motion of the soft mode $(l_s\approx 0.3$~\AA).
    The small splitting of each band at finite wave-vector along [100] is due to the Rashba mechanism, while the rigid shift at $\Gamma$ can be attributed to the two-phonon mechanism.  
    }
    \label{fig:DFT}
\end{figure}


To compute the second-order deformation potential of STO, relevant for the Ngai mechanism,  we deformed the tetragonal structure to simulate the presence of a Slater phonon displacement (Fig.~\ref{fig:Slater}), varying the displacement amplitude. 
We assume that the soft mode is well represented by the Slater mode as this is the dominant component found experimentally~\cite{Harada1970, Vogt1988N}. 
We parametrize the Slater mode amplitude by the relative displacement between the oxygen cage and the central titanium atom situated at site $\bm{ r}$,
\begin{equation}
    \bm{ u}_S(\bm{ r}) = \bm{ u}_{\rm O}(\bm{ r})  - \bm{ u}_{\rm Ti}(\bm{ r}). 
\end{equation} 

In the presence of the frozen mode, the three low-energy electronic bands displace upwards, as shown in Fig.~\ref{fig:DFT}. The $\Gamma$-point shift is quadratic in the Slater displacement as expected by symmetry.

We will consider only intraband contributions. So, for simplicity, we consider one band at a time and drop the band index. The band Hamiltonian reads,  
\begin{equation}\label{eq:hb}
    H_{b} = \sum_{\bm{ k}\sigma } \epsilon_{\bm{ k} }c_{\bm{ k}\sigma}^\dagger c_{\bm{ k}\sigma},
\end{equation}
where $\epsilon_{\bm{ k}}$ is the electron energy dispersion and $c_{\bm{ k} \sigma}$, $c_{\bm{ k} \sigma}^{\dagger}$ are fermionic operators which annihilate and create an electron with momentum $\bm{ k}$ and spin  $\sigma$. Furthermore, we neglect variations in band mass, etc., and assume a rigid band shift evaluated at $\Gamma$,   
\begin{equation}\label{eq:epshift}
   \epsilon_{\bm{ k} }= \epsilon_{0\bm{ k} }+\frac{1}{2}\sum_{mm'}\alpha_{mm'} u_{S,m}u_{S,m'}.
\end{equation}
Taking Cartesian components $m,m'$ coinciding with the 
crystallographic axes, only diagonal components of the 
$\alpha_{mm'}$ tensor are different from zero. Therefore we can define two independent parameters, $\alpha_{mm}=\alpha_\perp$ for displacements along the $c$-axis and  $\alpha_{mm}=\alpha_\parallel$ 
for displacements in the $a,b$-plane. 

We provide numerical estimates of $\alpha_\parallel$ and $\alpha_\perp$ in the next subsections. For empirical estimates based on experimental data, we will exploit the small tetragonality of the low-temperature structure, allowing us to neglect differences between basal and $c$-axis components and consider a single independent parameter $\alpha$. Without loss of generality, we will adopt the same simplification in the model computations.

 The coupled electron-lattice Hamiltonian with the density-two phonon interaction reads, 
\begin{equation}\label{eq:totham}
    H = H_{el} +H_{ph}+ H_{\textit{el-2ph}}
\end{equation}
The electronic Hamiltonian is assumed to be given by free electrons, 
\begin{equation}
    H_{el} = \sum_{\bm{ k}\sigma } \xi_{\bm{ k} }c_{\bm{ k}\sigma}^\dagger c_{\bm{ k}\sigma},
\end{equation}
where $\xi_{\bm{ k} }=\epsilon_{0\bm{ k}}-\mu$ and $\mu$ is the chemical potential.

 The phonon Hamiltonian reads, 
\begin{equation}\label{eq:hph}
    H_{ph}=\sum_{\bm{r}} \frac{\bm{p}_S^2}{2\mu_S} +\frac12 \sum_{\bm{r},\bm{r}'}
\bm{u}_S({\bm{r}})\cdot\overline{\bm{k}}
({\bm{r},\bm{r}'}) \cdot\bm{u}_S({\bm{r}'})+... 
\end{equation}
here $\overline{\bm{k}}$ is a spring-constant tensor,  $\bm{p}_S$ is the momentum conjugate to $\bm{u}_S({\bm{r}})$, 
 $\mu_S^{-1}=(3M_{\rm O})^{-1}+M_{\rm Ti}^{-1}$ is the reduced mass associated with the Slater mode and $M_{\rm O}$ and $M_{\rm Ti}$ are the oxygen and titanium masses. The ellipsis stands for anharmonic terms.

The electron-phonon interaction term is
\begin{equation}\label{eq:he2phr}
    H_{\textit{el-2ph}} =\frac12\alpha\sum_{\bm{ r}} \hat n({\bm{r}}) |\bm{u}_S({\bm{r}})|^2,
\end{equation}
where $\hat n({\bm{r}})=\sum_\sigma c_{\bm{ r}\sigma}^\dagger c_{\bm{ r}\sigma}$ is the density operator defined in terms of the real-space electron creation and annihilation operators associated with Wannier functions centered at $\bm r$. Here, for the sake of simplicity, we are assuming that the electron-two-phonon interaction is local.

 In the harmonic approximation, the phonon Hamiltonian can be written in quantized normal modes as
 \begin{equation}\label{eq:hphq}
    H_{ph}=\sum_{{\bm q}\lambda } \left(b_{{\bm q}\lambda}^\dagger b_{{\bm q}\lambda}+\frac12\right)\hbar\omega_{{\bm q}\lambda},
\end{equation}
where $b_{\bm{ q} \lambda}$, $b_{\bm{ q} \lambda}^{\dagger}$ are bosonic operators which destroy and create a phonon with momentum $\bm{ q}$ and polarization $\lambda$ and
$\omega_{\bm{ q}\lambda}$ is the phonon dispersion.

In terms of phonon operators, the electron-phonon interaction reads,
\begin{equation}\label{eq:he2ph}
\begin{split}
    H_{\textit{el-2ph}} & = \frac{1}{N}\sum_{\bm{ k}\sigma \bm{ q} \bm{ q}' \lambda \lambda '} 
    g_{\lambda \lambda' }(\bm{ q},\bm{ q} ') \\
    &\times 
    \bigg(b_{\bm{ q}\lambda}+b_{-\bm{ q} \lambda}^{\dagger}\bigg)\bigg(b_{\bm{ q}'\lambda'}+b_{-\bm{ q}' \lambda'}^{\dagger}\bigg) c_{\bm{ k}+\bm{ q}+\bm{ q}'\sigma}^\dagger c_{\bm{ k}\sigma}.\\
    \end{split}
\end{equation}
Here the coupling function $g_{\lambda \lambda' }(\bm{ q},\bm{ q} ')$ is equal to
\begin{equation}\label{eq:coup}
    g_{\lambda \lambda' }(\bm{ q},\bm{ q} ') =
    \frac{\hbar\alpha}{4\mu_S}\frac{1}{\sqrt{\omega_{\bm{ q}\lambda}\omega_{\bm{ q}'\lambda'}}}\hat{\bm{ n}}_{\bm{ q}\lambda} \cdot \hat{\bm{ n}}_{\bm{ q}'\lambda'},
\end{equation}
where 
$\hat{\bm{ n}}_{\bm{ q}\lambda}$ is the unit phonon polarization vector.

\begin{figure}[tb]
    \includegraphics[width=1\linewidth]{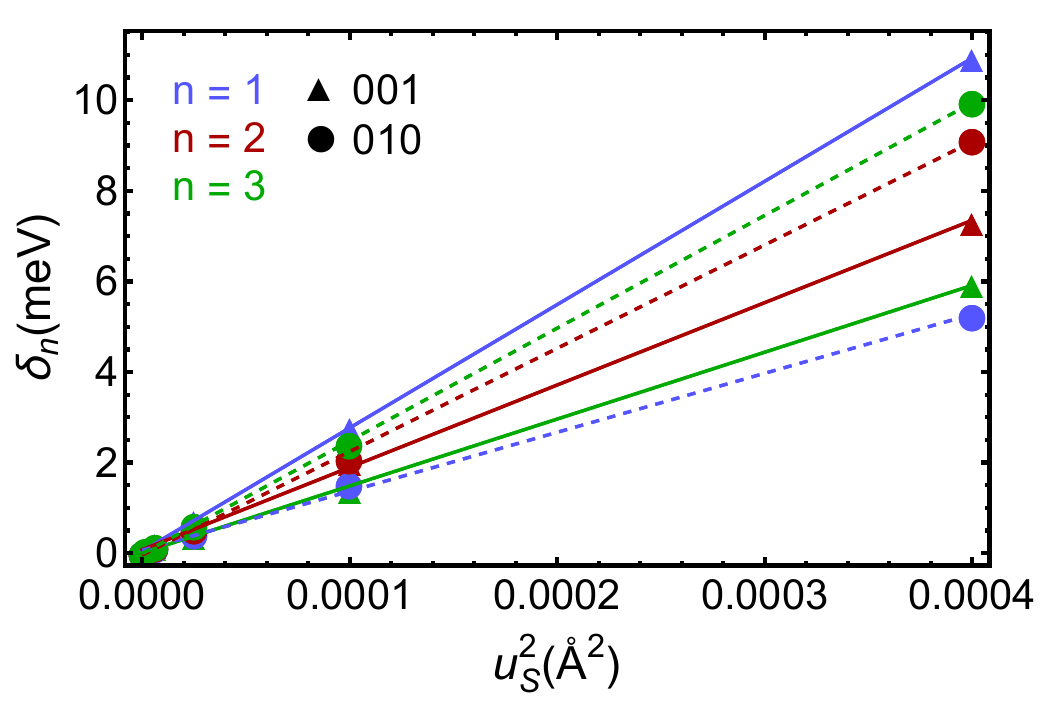}
    \caption{Band shift $\delta_n$ as a function of the relative displacement between oxygen and titanium squared $u_S^2$. Solid symbols are the DFT data, while the lines are a linear fit. The different colors indicate the three lower bands using the same conventions as Fig.~\ref{fig:DFT}  (band 1: blue, 2: red, and 3: green). Triangles and solid lines refer to computations for ${\bm u}_S$ along the $c$-direction, while circles and dashed lines are taken for ${\bm u}_S$ along (010) in pseudocubic coordinates ($a+b$ in the tetragonal ones), which corresponds nearly to the Ti-O-Ti bond direction in the basal plane.  
    }
\label{fig:BandShift}
\end{figure}

\subsection{Estimate of two-phonon deformation potential from eigenvalue shift}

To determine $\alpha_\perp$ and $\alpha_\parallel$ within DFT, we proceed as usual and interpret the Kohn-Sham eigenvalues as band energies. However, before proceeding, the DFT band shift has to be properly defined. Since adding a constant to the Kohn-Sham Hamiltonian leaves the density unchanged, Kohn-Sham eigenvalues are defined up to such a constant.  In order to have a physically meaningful band shift, we align the energies of the Sr $5s$ states in the different computations. In practice, we find that the VASP Sr $5s$ eigenvalues remain at the same energy in the presence of the lattice distortion (energy shifts less than 0.01 meV), and also the highest occupied state (belonging to the $2p$ O band) remains fixed with the same accuracy.    
We obtained the band structure shifts for $u_s=0.02~\text{\AA}$  and  1/2, 1/4, 1/8, and 1/16 of that value.  Figure~\ref{fig:BandShift} shows the band shifts at $\Gamma$ as a function of the squared displacement. The lines are linear fits, showing that the shift is quadratic in the displacement, as expected. 
From the fits, we obtained the results reported in Table~\ref{tab:alpha} for the three bands shown in Fig.~\ref{fig:DFT}, numbered from bottom to top. 

\begin{table}[tb]
   \caption{Second order deformation potential in eV/\AA$^2$ coefficients for out of plane      ($\alpha_{\perp}$) and inplane ($\alpha_{\parallel}$) displacements.   }
\begin{ruledtabular}
    \centering
    \begin{tabular}{c|ccc} 
  band  & 1 & 2 &3 \\
  \hline
$\alpha_{\perp}$ & 54.4  & 36.4  & 29.5  \\
$\alpha_{\parallel}$    &  30.1  & 45.6  & 49.8
    \end{tabular}
    \label{tab:alpha}
\end{ruledtabular}
\end{table}

Besides the Slater mode, also the  
'Last mode' (vibration of the cation out of phase with a rigid TiO$_3$) and the oxygen-octahedra distorting mode can contribute to the soft-mode~\cite{Harada1970,Vogt1988N}.
For these two modes, the $\alpha$ coefficients are in the range $10-40$~eV/\AA$^2$, but since their weights to the soft mode are small, we neglected them.
This is in contrast to the case of the  Rashba-like mechanism, where the oxygen-octahedron mode has a gigantic electron-phonon matrix element and can not be neglected~\cite{Gastiasoro2023}.
 
\subsection{Estimate of two-phonon deformation potential from total energies}
To cross-check the previous result for the second-order deformation potential $\alpha$, we performed an alternative estimate that relies only on total energy computations instead of the Kohn-Sham eigenvalues. This second estimate, therefore, does not depend on the issue of the zero of energy of the Kohn-Sham eigenvalues.

\begin{figure}[tb]
\centering
    \includegraphics[width=1\linewidth]{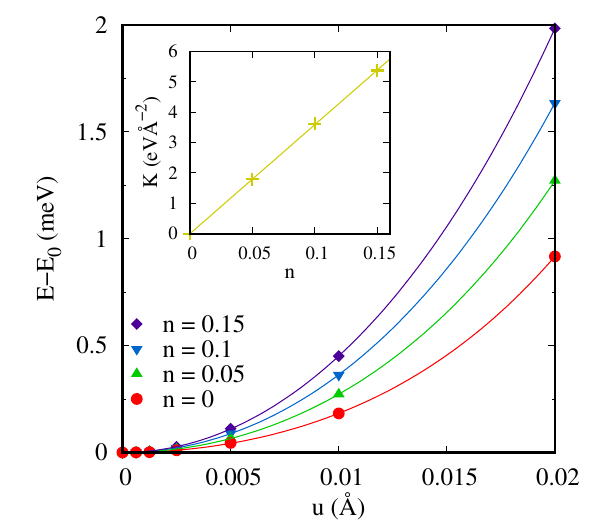}
    \caption{Total DFT energy as a function of the displacement in the Slater mode polarized along the tetragonal $a^T$ axis (pseudocubic [110] direction) for different band fillings $n$. Points are fitted using a quartic function $E(u)=Ku^2/2+V_4 u^4$, shown as a continuous line.
    The inset shows the effective stiffness constant as a function of band filling.}
    \label{fig:edudn}
\end{figure}

We base our second estimate on the hardening of the mode with doping. 
To obtain $\alpha$ we added a small amount  $n$ of electron charge per Ti to the lower band, compensated by a uniform positive background. For Brillouin-zone integrations we adopted the tetrahedron method with Bl\"ochl corrections~\cite{Blochl_prb1994} and a 34$\times$34$\times$24 k-point grid. From Eqs.~\eqref{eq:hph},\eqref{eq:he2phr}, it is clear that a uniform density renormalizes the spring constant for a uniform displacement of the Slater mode. Thus, we computed $\alpha$ from the total DFT energy as
\begin{equation}\label{eq:kdn}
 \alpha_{ij}=\frac{\partial^3 E_{DFT}}{\partial n \partial u_{S,i}  \partial u_{S,j} }.   
\end{equation}
Operationally, since only diagonal elements are non-zero,  we fit the total energy as a function of displacement with a fourth-order polynomial for each density, as shown in Fig.~\ref{fig:edudn},
and then obtain the linear term in the density dependence of this effective spring constant (inset). 
Using this method, we obtained for band one, $\alpha_\parallel=36$~eV/\AA$^2$, and $\alpha_\perp=57$~eV/\AA$^2$ (plot not shown), which is in fair agreement with the value found with the previous method.

Notice that rigorously in Eqs.~\eqref{eq:hph},\eqref{eq:he2phr} $\alpha$ is defined as the shift of the {\em bare} local energy. Instead, DFT takes into account exchange and correlation contributions to the total energy. Therefore, the present method assumes that contributions to the derivative in Eq.~\eqref{eq:kdn} from the exchange and correlation part of the energy are negligible.
The small discrepancy with the previous method is attributed to a contribution from those terms.

\subsection{Empirical estimates of the two-phonon deformation potential }

As mentioned in the introduction, the ferroelectric quantum critical point is characterized by the softening of 
the  $\Gamma$-frequency, $\omega_{\Gamma,\text{TO} }$, of two transverse optical modes (hereafter we will neglect the small difference of around 1 meV between the two phonon frequencies due to the tetragonal distortion). The phonon frequency is sensitive to carrier density~\cite{Bauerle1980N,Devreese2010,Edge2015,Saha2025,Fauque2025}.  
Analogous to our second theoretical computation, one can estimate $\alpha$ from the measured change of $\omega_{\Gamma,\text{TO} }$ with electron density. Assuming that the frequency follows the spring constant behavior in a quasiharmonic approximation, one obtains, 
$$\omega_{\Gamma,\text{TO} }^2(n)-\omega_{\Gamma,\text{TO} }^2(0)=\alpha n/\mu_S.$$

The room temperature mode frequency squared as a function of oxygen vacancy concentration was reported in Ref.~\cite{Bauerle1980N}. 
Assuming each vacancy corresponds to two conduction electrons, we deduce $\alpha=10$~eV/\AA$^2$. This, however, uses high-temperature data. Van der Marel and collaborators~\cite{Devreese2010} report the soft mode frequency in Nd-doped samples at $T=7$~K, and similar results were reported in recent neutron scattering experiments~\cite{Fauque2025}. 
Fitting Van der Marel's data and assuming each Nd provides one electron, we obtain $\alpha=21$~eV/\AA$^2$. 

In a subsequent paper
Van der Marel and collaborators~\cite{VanderMarel2019}, proposed an estimate of 
$\alpha$ 
based on spectral weight measurements and a simplified model of Ngai's mechanism. While this method is less direct, adopting their definitions, we deduce $\alpha=12$~eV/\AA$^2$, which is of the same order of magnitude as the estimates above. 

We attribute the discrepancy of these empirical estimates with the previous section's theoretical estimate to i) the quasiharmonic approximation, which is highly questionable close to the ferroelectric instability, and ii) the effect of chemical pressure induced by doping, which is also a relevant variable. For example, Ca doping~\cite{Bednorz1984} is known to induce ferrolectricity~\cite{Gastiasoro2020Review} without changing the carrier concentration. Clearly, doping-dependent strain will affect the phonon softening, making the estimate of $\alpha$ less accurate. 

We conclude by noting that
theoretical and experimental estimates differ by factors of  2 or 3, but the order of magnitude is the same. 
In the following, we take 
$\alpha= 40$ eV/\AA$^2$ obtained as the average of the theoretical values presented in Table~\ref{tab:alpha}.




\section{Toy-model and physical origin of the attraction}\label{toymod}
\begin{figure}[tb]
\centering
\includegraphics[width=1.0\linewidth]{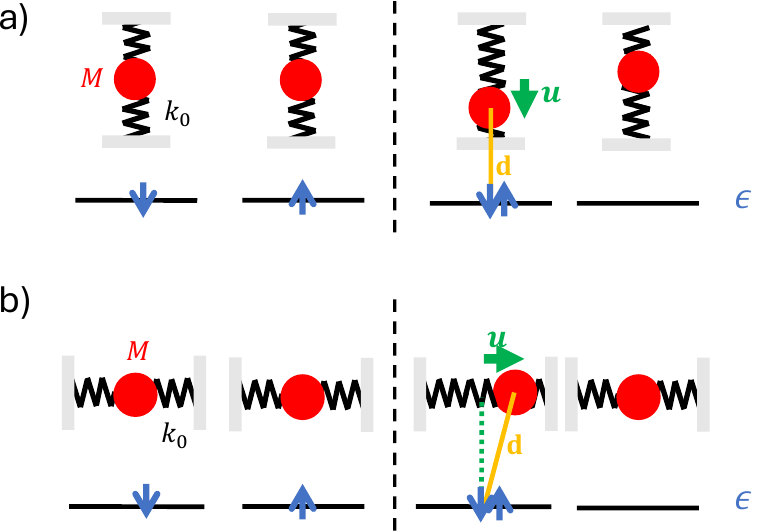}
    \caption{
    Linear (a) and quadratic (b) 
    Holstein models. An electronic level with single particle energy $\epsilon$ interacts with a harmonic oscillator representing an ion. Blue arrows represent the electrons. In (a), the distance of the ion to the electronic site varies linearly with the displacement, while in (b), it varies quadratically. Assuming that the energy of the electronic level is linear in the distance leads to the linear and quadratic models. Comparing the energy of two electrons in the same sites (left panels) with the one of two electrons in different sites (right) allows us to obtain the attractive interaction.  
    }
    \label{fig:Coupling}
\end{figure}

Before doing detailed computations, it is useful to obtain an intuitive understanding of the origin of the attraction in the Ngai mechanism. For this purpose, similarly to the discussion in Ref.~\cite{Han2024N}, we consider a standard (linear) 
Holstein model which will serve as a reference [Fig.~\ref{fig:Coupling}(a)], and a quadratic 
Holstein model [Fig.~\ref{fig:Coupling}(b)] both 
in the zero bandwidth or ``atomic" limit. 

In the standard Holstein model, at each site, the electron is coupled linearly to a harmonic oscillator representing the ionic vibrations.  The Hamiltonian  for one site reads, 
\begin{equation}\label{hamlincoup}
    \mathcal{H}_{1\text{p}}=\epsilon n + \frac{p^2}{2M}+\frac{1}{2}k_0u^2+\beta u n,
\end{equation}
where  $M$, $k_0$, $u$, and $p$ are the ion mass, the spring constant, the ion displacement, and the ion momentum, respectively. $n$ is the site electron number ($n=0,1,2$), $\beta$ is the electron-displacement coupling constant and 
$\epsilon$ is the bare diagonal energy of each site. 

In the second model, the coupling between electrons and ions is quadratic in the displacement, which can be schematized as an oscillator that moves tangentially with respect to the electronic site [Fig.~\ref{fig:Coupling}(b)]. In this case, the site Hamiltonian reads,  
\begin{equation}\label{hamquadcoup}
    \mathcal{H}_{2\text{p}}=\epsilon n + \frac{p^2}{2M}+\frac{1}{2}k_0u^2+\frac{1}{2}\alpha u^2 n.
\end{equation}

The two models are schematized in Fig.~\ref{fig:Coupling}. Notice that the coupling terms of the two Hamiltonians describe, respectively, the electron interaction with one and two phonons, and the parameter $\alpha$ of Eq.~\eqref{hamquadcoup} plays the same role as $\alpha$ in Eq.~\eqref{eq:he2phr}.


 
 Writing the Hamiltonian in terms of bosonic variables 
 and introducing the phonon number  $n_B$ we obtain the single-site energy. In the linear coupling case, the energy reads,
\begin{equation}\label{eq:holstein}
    E(n,n_B)=\epsilon n+\hbar\omega_0\biggl(n_b+\frac{1}{2}\biggr)-\frac{\beta^2n^2}{2M\omega_0^2}.
\end{equation}
with $\omega_0^2=k_0/M$.

Let us consider two sites and two electrons. 
We can define the effective electron-electron
interaction as the total energy difference between having  the electrons on different sites (left panels in Fig.~\ref{fig:Coupling}) and having both electrons on the same site (right panels),  
\begin{equation}\label{eq:veff}
V=E(2,0)+E(0,0)-2E(1,0).
\end{equation}
For $V<0$ there is attraction.
Using Eq.~\eqref{eq:holstein} we obtain the familiar result, 
\begin{equation}\label{effpotlin}
    V_{1\text{p}}=-\frac{\beta^2}{M\omega_0^2}
\end{equation}
which can be obtained directly from Eq.~\eqref{hamlincoup}, treating the displacement as a classical variable. 

In the case of quadratic coupling, the site energy using the quantized version of the energy is,  
\begin{equation}
    E(n,n_B)=\epsilon n+\hbar\omega_n\biggl(n_b+\frac{1}{2}\biggr),
\end{equation}
where
\begin{equation}
\omega_n=\omega_0\sqrt{1+\frac{\alpha n}{k_0}}.
\end{equation}
In this case, the phonon frequency depends on the occupancy, so the density dependence of the ground state energy is through a modification of the zero-point energy. Indeed, to leading order in $\alpha$ we obtain from Eq.~\eqref{eq:veff}, 
\begin{equation}\label{effpotquad}
    V_{2\text{p}}=-\frac{\hbar\alpha^2}{8M^2\omega_0^3}=-\frac{(\alpha l^2)^2}{2\hbar\omega_0},
\end{equation}
In contrast with Eq.~\eqref{effpotlin}, which can be obtained classically, the attraction here is manifestly a purely quantum mechanical effect since $\hbar$ appears as a prefactor.  In the second form, we use the harmonic oscillator length $l^2\equiv\hbar/(2M\omega_0)$.

The linear and quadratic cases lead to attraction but for quite different reasons. An important consequence is that in the linear model, Eq.~\eqref{effpotlin}, the attraction scales as $1/\omega_0^2$. This coincides with the finding in BCS theory based on the standard linear electron-phonon coupling. Instead, for the quadratic model, it scales as $1/\omega_0^3$. 
This suggests that the quadratic model can benefit more from soft-phonon effects as found near the FE instability. Indeed,
making the replacements $\hbar\omega_0\rightarrow \hbar\omega_{\Gamma,\text{TO} }\approx 1 $ meV (valid for undoped STO~\cite{Gastiasoro2020Review}), $l\rightarrow l_s\approx 0.3$~\AA, and $\alpha=40$~eV/\AA$^2$ one gets $\alpha l^2\approx 3.6$~eV and $V_{2\text{p}}\approx -6000$~eV, a surprisingly large value. 
Taking the more realistic value $\omega_{\Gamma,\text{TO} }\approx6$~meV measured at 0.9~\% doping \cite{Devreese2010} and an
experimental $\alpha\approx~20$~eV/\AA$^2$ yields $V_{2\text{p}}\approx -8$ eV.
It is still a huge value, much larger than the bandwidth. This suggests that the two-phonon mechanism can be very efficient in producing pairing and perhaps superconductivity. 
Instead, we will show below that, in a more realistic model, the attraction is substantially reduced. The gross overestimation found here can be traced to neglecting the phonon dispersion.

\section{Effective two-phonon mediated electron-electron interaction}\label{Eff}

Since only transverse modes become soft, we 
restrict the two sums over phonon polarizations in Eq.~\eqref{eq:he2ph} to the transverse optical modes $\lambda =$(TO1,TO2) \cite{Muzzi}. 
For simplicity, we assume the system is isotropic so the optical transverse branches are degenerate with energy $\omega_{\bm{ q} \text{TO}1}=\omega_{\bm{ q} \text{TO}2}=\omega_{\bm{ q} \text{TO}}$.

The following dispersion reproduces the experimental behavior of the dispersion at low energies \cite{MNG2} and will be used in the computations: 
\begin{equation}\label{PhononDispersion}
     \omega_{\bm{ q}\text{TO}} = \sqrt{\omega_{\Gamma,\text{TO} }^2 + c^2 \bm{ q}^2},
\end{equation}
where $c$ is the velocity of the transverse mode at the ferroelectric critical point given by $\omega_{\Gamma,\text{TO} }=0$, defining the acoustic-dispersion regime. 
We will use $\omega_{\Gamma,\text{TO} }$ as a free parameter that controls the distance to the ferroelectric quantum critical point.  

To compute the effective attraction, we proceed similarly to the linear electron-phonon coupling~\cite{Bardeen1955}. We 
treat   $H_{\textit{el-2ph}}$ in Eq.~\eqref{eq:he2ph} as a perturbation and perform a Schrieffer-Wolff transformation to eliminate it to first order. The
total Hamiltonian $H$ [Eq.~\eqref{eq:totham}] is mapped
into a new one with the same eigenvalues through a canonical transformation,
\begin{equation}\label{eq:htilde}
    \Tilde{H}=e^{-S}He^S=H+[H,S]+\frac12[[H,S],S]+...
\end{equation}
where $S$ is an anti-hermitian operator, i.e. $S^\dagger=-S$.
We chose $S$ such that 
\begin{equation}\label{eq:scon}
    [H_{el}+H_{ph},S]=-H_{el-2ph}
\end{equation}
so that the electron-two-phonon interaction is canceled to first order, and the transformed Hamiltonian becomes,
\begin{equation}\label{eq:htilde2}
    \Tilde{H}=H_{el}+H_{ph}+\frac12[H_{el-2ph},S]+...
\end{equation}
Here, $S$ is given by, 
\begin{equation}
 \begin{split}
S&=\sum_{{\bm k}{\bm q}{\bm q}'\sigma}g_{\lambda\lambda'}({\bm q},{\bm q}')c_{{\bm k}+{\bm q}+{\bm q}'\sigma}^\dagger c_{{\bm k}\sigma} (
 x_{{\bm k}{\bm q}{\bm q}'\lambda\lambda'}b_{-{\bm q}\lambda}^\dagger b_{-{\bm q}'\lambda'}^\dagger\\
&
+ y_{{\bm k}{\bm q}{\bm q}'\lambda\lambda' }b_{-{\bm q}\lambda}^\dagger b_{{\bm q}'\lambda'}
+z_{{\bm k}{\bm q}{\bm q}'\lambda\lambda'}b_{{\bm q}\lambda}b_{{\bm q}'\lambda'}), 
 \end{split}
\end{equation}
and the coefficients $x_{{\bm k}{\bm q}{\bm q}'},y_{{\bm k}{\bm q}{\bm q}'}$, and $z_{{\bm k}{\bm q}{\bm q}'}$ are obtained by imposing Eq.~\eqref{eq:scon}.
The last term in Eq.~\eqref{eq:htilde2} yields the effective interaction plus bosonic terms, which are neglected. Also, higher-order terms in the expansion are neglected. 
To leading order in the electron-phonon matrix element, the effective interaction reads,  
\begin{equation}\label{eq:Htil}\begin{split}
    &\tilde{H}_{\text{el-el}} = \frac{1}{2N^2}\\
  &\times \sum_{\bm{ k} \bm{ k} ' \sigma \sigma ' \bm{ q} \bm{ q} '}
 \mathcal{V}_\text{eff}({\bm k},\bm{ k} ',{\bm q},{\bm q}') 
     c_{\bm{ k}+\bm{ q} + \bm{ q} '\sigma}^{\dagger}c_{\bm{ k} '- \bm{ q} - \bm{ q} ' \sigma '}^{\dagger}c_{\bm{ k}' \sigma '}c_{\bm{ k}\sigma},
 \end{split}\end{equation}
with 
\begin{equation}\label{InterazEff}
   \begin{split}
& \mathcal{V}_\text{eff}({\bm k},{\bm k}',{\bm q},{\bm q}')   \\ 
&= \abs{g(\bm{ q},\bm{ q} ')}^2
\frac{4\hbar( \omega_{\bm{ q}\text{TO}} + \omega_{\bm{ q}' \text{TO} } )}{(\epsilon_{\bm{ k}} -\epsilon_{\bm{ k} + \bm{q} + \bm{ q}' })^2-\hbar^2(\omega_{\bm{ q} \text{TO} } + \omega_{\bm{ q}' \text{TO} })^2}, 
    \end{split}
\end{equation}
where we defined the total coupling constant to transverse modes,
\begin{equation}\label{CouplingFunction2}
\begin{split}
{g(\bm{ q},\bm{ q} ')} & \equiv \sum_{\lambda\lambda'}{g_{\lambda \lambda' }(\bm{ q},\bm{ q} ')}\\
&={\frac{\hbar\alpha}{4 \mu_S
\sqrt{\omega_{\bm{ q} \text{TO} }\omega_{\bm{ q}' \text{TO} }}}}
f^{\frac{1}{2}}(\hat{\bm{ q}}, \hat{\bm{ q}} '), 
\end{split}
\end{equation}
The angular function $f(\hat{\bm{ q}}, \hat{\bm{ q}} ')\equiv 1 + (\hat{\bm{ q}} \cdot \hat{\bm{ q}} ')^2$ results from the scalar product in Eq.~\eqref{eq:coup}. Because we assume an isotropic system, the two polarization vectors in Eq.~\eqref{eq:coup} can be taken as two mutually orthogonal vectors and orthogonal to the longitudinal direction $\hat{\bm{ q}}$, which leads to the expression for $f$. 

The expression for the effective interaction, Eq.~\eqref{InterazEff}, is similar to the textbook expression for the one-phonon case, except that the sum of the two phonon frequencies appears in the place of a single phonon frequency and two sums over phonon momentum instead of one. We have the usual situation that the interaction is attractive  when
\begin{equation}
    \abs{\epsilon_{\bm{ k}}-\epsilon_{\bm{ k} + \bm{ q} + \bm{ q}' }}<\omega_{\bm{ q} \text{TO} } + \omega_{\bm{ q}' \text{TO} }.
\end{equation}
Therefore, one expects a Cooper instability due to the exchange of two phonons. 

As a check, we can set the two fermionic energies to be the same and consider the case of a dispersionless phonon, which yields a result analogous to the toy model [Eq.~\eqref{effpotquad}] apart from angular dependencies, 
\begin{equation}
    \mathcal{V}_\text{eff}({\bm k},\bm{ k}, {\bm q},{\bm q'})=-\frac{\hbar\alpha^2}{8\mu_S^2\omega_{\Gamma,\text{TO} }^3} f(\hat{\bm{ q}}, \hat{\bm{ q}} ').
\end{equation}
Clearly, the physical origin of the attractive interaction is the same as in the toy model, namely the renormalization of the zero-point phonon energy.  The denominators in Eqs.~\eqref{InterazEff}-\eqref{CouplingFunction2} involve phonons at all momenta, and only in a very small region of phase space the phonon softening becomes relevant. As we shall see, this strongly penalizes the two-phonon attraction as a mechanism for superconductivity in STO.

\section{One-loop Eliashberg Equations}
The effective interaction Eq.~\eqref{InterazEff} has a strong momentum and energy dependence, which becomes a frequency dependence in a diagrammatic formalism. Therefore, we will use Eliashberg equations to estimate $T_c$. In the following, we set $\hbar=k_B=1$ and will restore them when they become useful for clarity. 

Eliashberg equations neglect vertex corrections, which are usually justified by applying the Migdal theorem~\cite{ALLEN19831,Marsiglio_2020}. This requires that bosonic frequencies (in our case, twice a bosonic frequency) are smaller than the Fermi energy. 
We model the phonon dispersion with Eq.~\eqref{PhononDispersion}, where 
the parameter $\hbar c/a= 13 $~meV is taken from neutron scattering experiments~\cite{Shirane1969N}. Here  
 $a=3.9$~\AA~ is the lattice constant assuming a cubic phase ($a\approx c^T/2$). Typical values~\cite{Lin2014} of the Fermi energy in STO are in the range $1\sim8$~meV. 
Therefore, while the frequency around $\Gamma$ can be close to the adiabatic limit, for zone boundary phonons ($q\sim \pi/a$) the Migdal theorem does not apply, and vertex corrections may be important~\cite{Pietronero1995,Grimaldi1995,Cappelluti2002,NuovoEliash}. On the other hand, the one-loop approximation is still well justified in the weak coupling limit. It will be clear below that this is the case for the two-phonon mechanism in STO, and therefore, we can neglect vertex corrections. 

The solution of the one-loop Eliashberg equation under the present condition is not standard. Because of this, we present the full derivation with some detail. 

We introduce the Nambu spinors evaluated at the imaginary time $\tau$, where $-\beta\le\tau\le\beta$ and $\beta=1/T$ is the Boltzman factor,
\begin{equation}
    \Psi_{\bm{ k}}(\tau)=
    \begin{pmatrix}
 c_{\bm{ k}\uparrow}(\tau)\\
c_{-\bm{ k}\downarrow}^\dagger(\tau)
\end{pmatrix},
\quad
\Psi_{\bm{ k}}^\dagger(\tau)=
    \begin{pmatrix}
 c_{\bm{ k}\uparrow}^\dagger(\tau) & c_{-\bm{ k}\downarrow}(\tau)
\end{pmatrix},
\end{equation}
where $c_{\bm{ k}\sigma}(\tau)$, $c_{\bm{ k}\sigma}^\dagger(\tau)$ are the fermion annihilation and creation operators in the interacting picture. The finite-temperature Matsubara-Green's function reads 
\begin{equation}\label{NambuMatsubaraGreen}
\begin{split}  
   &\hat{\mathcal{G}}(\bm{ k},\tau) = -\langle T_\tau \Psi_{\bm{ k}}(\tau)\Psi_{\bm{ k}}^{\dagger}(0) \rangle = 
   \begin{pmatrix}
       \mathcal{G}(\bm{ k},\tau) & \mathcal{F}(\bm{ k},\tau) \\
       \mathcal{F}^*(\bm{ k},\tau) & -\mathcal{G}(-\bm{ k},-\tau)
   \end{pmatrix}.\\
   \end{split}
\end{equation}
Here, the normal state component, $\mathcal{G}(\bm{ k},\tau)
=- \langle T_\tau c_{\bm{ k}\uparrow}(\tau) c_{\bm{ k}\uparrow}^{\dagger}(0)\rangle$, describe single-particle excitations, while the Gor'kov's functions in the anomalous channel,  $\mathcal{F}(\bm{ k},\tau)=-\langle T_\tau c_{\bm{ k}\uparrow}(\tau) c_{-\bm{ k}\downarrow}(0)\rangle$, describe the Cooper pairs amplitude \cite{ALLEN19831}. Because time-reversal invariance is preserved in our theory, spin indices can be omitted. We also introduce the Green functions in the imaginary frequency axis,
\begin{equation}
   \hat{\mathcal{G}}(\bm{ k},i\omega_{n}) =
    \begin{pmatrix}
        \mathcal{G}(\bm{ k},i\omega_n) & \mathcal{F}(\bm{ k},i\omega_n) \\
       \mathcal{F}^*(\bm{ k},i\omega_n) & -\mathcal{G}(-\bm{ k},-i\omega_n)
    \end{pmatrix},
  \end{equation}
where the Fourier transform is defined as, 
\begin{equation}
    \hat{\mathcal{G}}(\bm{ k},i\omega_n)=\frac{1}{2}\int_{-\beta}^{\beta} d\tau e^{i\omega_n \tau}\hat{\mathcal{G}}(\bm{ k},\tau),
\end{equation}
and $\omega_{n}=(2n+1)\pi T$ is the fermionic Matsubara frequency, with $n\in\mathbb{Z}$.
  
The Matsubara Green's function is the solution of the Dyson equation,
\begin{equation}\label{DysonEq}
    \hat{\mathcal{G}}^{-1}(\bm{ k},i\omega_n)=\hat{\mathcal{G}}_{0}^{-1}(\bm{ k},i\omega_n)-\hat{\Sigma}(\bm{ k},i\omega_n).
\end{equation}
with the unperturbed Green's function defined as
\begin{equation}
    \hat{\mathcal{G}}_{0}^{-1}(\bm{ k},i\omega_n) = i\omega_n\hat{\tau}_0-\xi_k\hat{\tau}_3.
\end{equation}
Here $\hat{\tau}_0$ is the identity matrix and $\hat{\tau}_i$ ($i=1,2,3$) are Pauli matrices. 

The second term appearing in Eq.~\eqref{DysonEq} is the $2\times2$ electron self-energy, which, in its most general form, is given by
\begin{equation}
    \begin{split}    
    \hat{\Sigma}(\bm{ k},i\omega_n) & = i\omega_n(1-Z(\bm{ k},i\omega_n))\hat{\tau}_0+\chi(\bm{ k},i\omega_n)\hat{\tau}_3\\
    &+\phi(\bm{ k},i\omega_n)\hat{\tau}_1+\Bar{\phi}(\bm{ k},i\omega_n)\hat{\tau}_2.\\
    \end{split}
    \label{self-energy}
\end{equation}
In this expression, there are four unknown quantities (to be fixed by solving the Eliashberg equation): two diagonal terms, i.e. those multiplying $\hat{\tau}_0$ and $\hat{\tau}_3$, $Z(\bm{ k},i\omega_n)$ and $\chi(\bm{ k},i\omega_n)$,
and two off-diagonal terms $\phi(\bm{ k},i\omega_n)$ and $\bar{\phi}(\bm{ k},i\omega_n)$ composing the order parameter (i.e. the gap function) \cite{ALLEN19831,Marsiglio_2020}. 
Since ${\phi}(\bm{ k},i\omega_n)$ and $\Bar{\phi}(\bm{ k},i\omega_n)$ differ by a phase factor and we have time reversal symmetry, it is possible to choose a gauge where $\Bar{\phi}(\bm{ k},i\omega_n)=0$ \cite{ALLEN19831}.

Using Eq.~\eqref{self-energy} and \eqref{DysonEq} it is possible to find the Matsubara-Green's function in the Nambu formalism 
\begin{equation}\label{DefGreen}
    \begin{split}    
 &   \hat{\mathcal{G}}(\bm{ k},i\omega_n) 
    = -\frac{1}{\Theta(\bm{ k},i\omega_n)} [i\omega_nZ(\bm{ k},i\omega_n)\hat{\tau}_0\\    
    & +
    (\xi_{\bm{ k}}+\chi(\bm{ k},i\omega_n))\hat{\tau}_3 +\phi(\bm{ k},i\omega_n)\hat{\tau}_1 ],
    \end{split}
\end{equation}
where
\begin{equation}
\begin{split}
    \Theta(\bm{ k},i\omega_n) & =-(i\omega_nZ(\bm{ k},i\omega_n))^2+(\xi_{\bm{ k}}+\chi(\bm{ k},i\omega_n))^2\\    &+\phi^2(\bm{ k},i\omega_n).
    \end{split}
\end{equation}
The gap function is defined as
\begin{equation}
   \Delta(\bm{ k},i\omega_n)=\frac{\phi(\bm{ k},i\omega_n)}{Z(\bm{ k},i\omega_n)}.
\end{equation}

\subsection{Equations for the Conventional electron-phonon interaction}

To fix notations and clarify the formalism, we start by recalling the derivation of  Eliashberg equations in the familiar case of a conventional one-phonon interaction. 
The Hamiltonian reads, 
\begin{equation}
    H_\textit{el-ph}=\frac{1}{\sqrt{N}}\sum_{\bm{ k}\bm{ k}'}
    g^{1ph}_{\bm{ k}-\bm{ k}'} A_{\bm{ k}-\bm{ k}'}
    \Psi_{\bm{ k}'}^\dagger\hat{\tau}_3\Psi_{\bm{ k}},
\end{equation}
where $N$ is the lattice-sites number, $g^{1ph}_{\bm{ k}-\bm{ k}'}$ is the coupling function and we defined $A_{\bm{ q}}\equiv b_{\bm{ q}}+b_{-\bm{ q}}^\dagger$.

Keeping only the one-loop approximation to the electronic self-energy leads to, 
\begin{equation}\label{FirstOrderSigma}
\begin{split}
    &\hat{\Sigma}(\bm{ k},i\omega_n)=-\frac{T}{N}\\
    &\times\sum_{\bm{ k}'i\omega_{n'}}\abs{g^{1ph}_{\bm{ k}-\bm{ k}'}}^2D(\bm{ k}-\bm{ k}',i\omega_{n}-i\omega_{n'})\hat{\tau}_3\hat{\mathcal{G}}(\bm{ k}',i\omega_{n'})\hat{\tau}_3,
    \end{split}
\end{equation}
where 
\begin{equation}
    D(\bm{ q},i\Omega_{n})=\int_{0}^{\infty}  B(\bm{ q},\omega)\frac{2\omega }{(i\Omega_{n})^{2}-\omega^{2}} d\omega ,
\end{equation}
is the phonon propagator, $i\Omega_n=2n\pi T$ is a bosonic Matsubara frequency and   $B(\bm{ q},\omega)$
is the electron-phonon spectral function. Here,
the momentum-resolved Eliashberg function is defined as \cite{NuovoEliash} 
\begin{equation}
    \alpha^2 F(\bm{ q},\omega) = \abs{g^{1ph}_{\bm{ q}}}^2B(\bm{ q},\omega),
\end{equation}
and the effective electron-electron interaction reads~\cite{NuovoEliash} 
\begin{equation}\label{DefVeffEliash}
\begin{split}
    &{V_\text{eff}(\bm{ q},  i\Omega_n)} =\abs{g^{1ph}_{\bm{ q}}}^2D(\bm{ q},i\Omega_{n})\\
    &=\int_{0}^{\infty} \alpha^2 F(\bm{ q},\omega) \frac{2\omega}{(i\Omega_{n})^{2}-\omega^{2}} d\omega .
\end{split}
\end{equation}
Combining Eq.~\eqref{DefGreen} and Eq.~\eqref{FirstOrderSigma} yields a self-consistent matrix equation that corresponds to the imaginary axis Eliashberg equations 
\begin{equation}\label{eleq1}
 \begin{split}
    &Z(\bm{ k},i\omega_n)-1\\
    &=-\frac{T}{N\omega_n}\sum_{\bm{ k}'i\omega_{n'}}\frac{{V_\text{eff}(\bm{ k}-\bm{ k}', i\omega_n-i\omega_{n'})}}{\Theta(\bm{ k}',i\omega_{n'})}\omega_{n'}Z(\bm{ k}',i\omega_{n'}),
    \end{split} 
\end{equation}
\begin{equation}\label{eleq2}
 \begin{split}
&      Z(\bm{ k},i\omega_n)\Delta(\bm{ k},i\omega_n) =-\frac{T}{N}
\\
      &\times \sum_{\bm{ k}'i\omega_{n'}}\frac{{V_\text{eff}(\bm{ k}-\bm{ k}', i\omega_n-i\omega_{n'})}}{\Theta(\bm{ k}',i\omega_{n'})} Z(\bm{ k}',i\omega_{n'})\Delta(\bm{ k}',i\omega_{n'}),
      \end{split}  
\end{equation}
\begin{equation}\label{eleq3}
\begin{split}
   & \chi(\bm{ k},i\omega_n)=\frac{T}{N}
\\
      &\times 
    \sum_{\bm{ k}'i\omega_{n'}}\frac{{V_\text{eff}(\bm{ k}-\bm{ k}', i\omega_n-i\omega_{n'})}}{\Theta(\bm{ k}',i\omega_{n'})}(\xi_{\bm{ k}'}+\chi(\bm{ k}',i\omega_{n'})). \end{split}  
\end{equation}
With these equations, it is possible to determine the behavior of the gap function and the critical temperature $T_c$, including the dynamical effects of phonons. The imaginary axis Eliashberg equations provide real solutions for $Z(\bm{ k},i\omega_{n})$, $\Delta(\bm{ k},i\omega_{n})$ and $\chi(\bm{ k},i\omega_{n})$ \cite{ALLEN19831}.

\subsection{Equations for the two-phonon-electron interaction}
The Hamiltonian for the two-phonon 
Eq.~\eqref{eq:he2ph} in the Nambu formalism reads, 
\begin{equation}
  {H}_{\text{el-2ph}} = \frac{1}{N}\sum_{\bm{ k} \sigma \bm{ q} \bm{ q}' \lambda \lambda '} 
    g_{\lambda \lambda'}(\bm{ q},\bm{ q} ')
    A_{\bm{ q} \lambda , \bm{ q}' \lambda'}\Psi_{\bm{ k}+\bm{ q}+\bm{ q}'}^\dagger\hat{\tau}_3\Psi_{\bm{ k}},
\end{equation}
where we defined
\begin{equation}
A_{\bm{ q} \lambda,\bm{ q}' \lambda'}\equiv \bigg(b_{\bm{ q}\lambda}+b_{-\bm{ q}\lambda}^{\dagger}\bigg)\bigg(b_{\bm{ q}'\lambda'}+b_{-\bm{ q}'\lambda'}^{\dagger}\bigg).
\end{equation}

The two-phonon propagator is defined as
\begin{equation}\begin{split} 
 &D_{\lambda\lambda'}(\bm{ q},\bm{ q}',i\Omega_n)\\
 &=-\int_{0}^{\beta} d\tau e^{i\Omega_n \tau}\langle T_\tau A_{\bm{ q}\lambda,\bm{ q}'\lambda'}(\tau)A_{\bm{ q}\lambda,\bm{ q}'\lambda'}^{\dagger}(0) \rangle.
\end{split}\end{equation}
As in  Sec. \ref{Eff}, we assume that the system is isotropic and the optical transverse branches are degenerate ($\omega_{\bm{ q}\text{TO1}}=\omega_{\bm{ q} \text{TO2}}=\omega_{\bm{ q} \text{TO} }$). Due to this isotropy, the two-phonon Green's function $D_{\lambda\lambda}(\bm{ q},\bm{ q}',i\omega_n)$ does not depend on polarization and the index $\lambda$ can be dropped, yielding, 
\begin{equation}
     D(\bm{ q},\bm{ q}', i\Omega_n) =\frac{4(\omega_{\bm{ q}}+\omega_{\bm{ q}'})}{(i\Omega_{n})^{2}-(\omega_{\bm{ q}}+\omega_{\bm{ q}'})^{2}}.\label{eq:D2ph}
\end{equation}
Similarly to the case of the conventional electron-phonon interaction, we can define the effective two-phonon mediated interaction as a function of the total phonon momentum $\bm{ Q}=\bm{ q}+\bm{ q}'$ as, 
\begin{equation}\label{eq:veff2ph}
\begin{split} 
&V_\text{eff}\big(\bm{ Q}, i\Omega_n\big)=\\
&=\frac{1}{N}\sum_{\bm{ p}}\abs{g \bigg({\frac{\bm{ Q}}{2}+\bm{ p},\frac{\bm{ Q}}{2}-\bm{ p} }\bigg)}^2 D\bigg(\frac{\bm{ Q}}{2}+\bm{ p}, \frac{\bm{ Q}}{2}-\bm{ p},i\Omega_n\bigg),\end{split}
\end{equation}
where $\bm{ p}=(\bm{ q}-\bm{ q}')/2$ and the coupling function ${g(\bm{ q},\bm{ q} ')}$ contains the sums over polarizations [Eq.~\eqref{CouplingFunction2}].  
With these definitions, the Eliashberg equations [Eqs.~\eqref{eleq1}-\eqref{eleq3}] also apply to the two-phonon mediated effective interaction defined in Eq.~\eqref{eq:veff2ph}.
Equations \eqref{eq:D2ph}- \eqref{eq:veff2ph} provide the dynamical (retarded) generalization of the effective interaction of  Eq.~\eqref{InterazEff}. 

Similar to Eq.~\eqref{DefVeffEliash} we can express $V_\text{eff}\big(\bm{ Q},i\omega_n \big)$ as an integral over the frequency $\Omega$
\begin{equation}\label{Veffica}
{V_\text{eff}\big(\bm{ Q}, i\Omega_n \big)} =
    \int_{0}^{\infty} \alpha^2 F(\bm{ Q},\Omega) \frac{2\Omega}{(i\Omega_{n})^{2}-\Omega^{2}}  d\Omega
\end{equation}
where
\begin{equation}\label{eliashbergfnc}
\begin{split} 
 &   \alpha^2 F(\bm{ Q},\Omega) =\\ &=\frac{2}{N}\sum_{\bm{ p}}\abs{g \Bigg({\frac{\bm{ Q}}{2}+\bm{ p},\frac{\bm{ Q}}{2}-\bm{ p}}\Bigg)}^2 \delta \big(\Omega - \omega_{\frac{\bm{ Q}}{2}+\bm{ p}}^{\text{TO}} -\omega_{\frac{\bm{ Q}}{2}-\bm{ p}}^{\text{TO}}\big)
\end{split}\end{equation}
is defined as the Eliashberg function in the case of the two-phonon interaction.

\subsection{Momentum-dependent cutoff for two-phonon scattering}\label{sec:wd}
To analytically compute the Eliashberg function, we consider a continuum model where the soft phonon dispersion is given by Eq.~\eqref{PhononDispersion}.
To proceed, we calculated the two-phonon density of states (DOS) defined as
\begin{equation}
\begin{split} 
   {\cal D}(\bm{ Q},\Omega) =& \frac{V}{N}\int \frac{d^{3}p}{(2\pi)^3}\delta \big(\Omega - \omega_{\frac{\bm{ Q}}{2}+\bm{ p}} -\omega_{\frac{\bm{ Q}}{2}-\bm{ p}}\big),\\
   &\quad \quad\quad \quad \quad \quad \text{ when } \Omega<2\omega_D(\bm{ Q}),
\end{split}    
\end{equation}
and,
\begin{equation}
   {\cal D}(\bm{ Q},\Omega) = 0,\quad \quad \quad \quad\quad \quad \text{otherwise.}  
\end{equation}
Here, in the spirit of the Debye model, we introduced a momentum-dependent frequency cutoff,  $\omega_D({\bm Q})$, to limit the number of normal modes in the continuum. $\omega_D({\bm Q})$ is defined by the condition, 
\begin{equation}\label{intomegad1}
    \int_0^{\infty} {\cal D}(\bm{ Q},\Omega)d\Omega=\int_0^{2\omega_D(\bm{ Q})} {\cal D}(\bm{ Q},\Omega)d\Omega=1.
\end{equation}
The factor of 2 ensures that for $Q=0$, $\omega_D$  coincides with the usual Debye frequency in the acoustic-dispersion regime ($\omega_{\Gamma,\text{TO} }=0$),
\begin{equation}\label{OmegaD0}
    \omega_{D \Gamma} \equiv \omega_D({\bm 0})= cq_D 
\end{equation}
with $q_D=(6\pi^2)^{1/3}/a$.
Taking $\hbar c/a=13 $~meV \cite{Shirane1969N}
yields $\hbar\omega_{D \Gamma}=50$~meV for STO at the ferroelectric critical point. Away from that, small values of $\omega_{\Gamma,\text{TO} }$ introduce a correction to $\hbar\omega_{D \Gamma}$ of order  $\omega_{\Gamma,\text{TO} }/\omega_{D \Gamma}^2$, which 
will be neglected in the computations below. 

Notice that the Debye frequency model, introduced here for solving the Eliashberg equations,  should not be confused with the usual Debye frequency of STO, as we are taking into account only the two TO modes and not the whole phonon spectra. 

Computing the two-phonon density of states and solving Eq.~\eqref{intomegad1}, one can find an analytic solution for $\omega_D(\bm{ Q})$. However, such an expression is too cumbersome to be used. We find that the following approximated expression
\begin{equation}\label{approxdeb}
    \omega_D(\bm{ Q})=c\sqrt{\bigg(\frac{Q}{2}\bigg)^2+\bigg(\frac{\omega_{D\Gamma}}{c}\bigg)^2},
\end{equation}
reproduces the exact momentum-Debye frequency with good accuracy, and it will be used in the following.   

\subsection{Two-phonon  Eliashberg function and effective interaction}
The cutoff frequency introduced above implies that
\begin{equation}
    \alpha^2 F(\bm{ Q},\Omega)=0 \quad\text{when}\quad \Omega>2\omega_D(\bm{ Q}).
\end{equation}

Unfortunately, the factor
$f(\hat{\bm{ q}}, \hat{\bm{ q}} ')$ in Eq.~\eqref{CouplingFunction2} makes the Eliashberg function Eq.~\eqref{eliashbergfnc} and the effective potential integral Eq.~\eqref{Veffica} analytically intractable. However, since the function $f$ has lower and upper bounds
($1\le{f}\le2$) we can replace the function by a constant value $\Bar{f}$, where $1\le\Bar{f}\le2$, which will yield upper and lower bounds for $T_c$. 

With these approximations 
Eq.~\eqref{eliashbergfnc} can be computed analytically, yielding, 
\begin{equation}\label{eq:alfaexact}
 \begin{split}
   \alpha^2 F(\bm{ Q},\Omega)=&\frac{V_0}{2}\sqrt{\frac{{\Omega^2}-c^2{Q}^2-4{\omega_{\Gamma,\text{TO} }^2}}{{\Omega^2}-c^2Q^2}} \\
   &\times\Theta\big(\Omega-2\omega_{{\bm{ Q}}/{2}}\big)\Theta\big(2\omega_D({\bm{ Q}})-\Omega\big), 
 \end{split}
\end{equation}
where we defined 
\begin{equation}\label{V0}
    V_0\equiv\frac{3\hbar\alpha^2}{8\mu_S^2
    }
    \frac{\bar{f}}{{\omega_{D\Gamma}}^3}.
\end{equation}
As it will become clear below, $V_0$  parametrizes the strength of the effective attractive interaction. 
\begin{figure}[tb]
 \centering
  \includegraphics[width=1.07\linewidth]{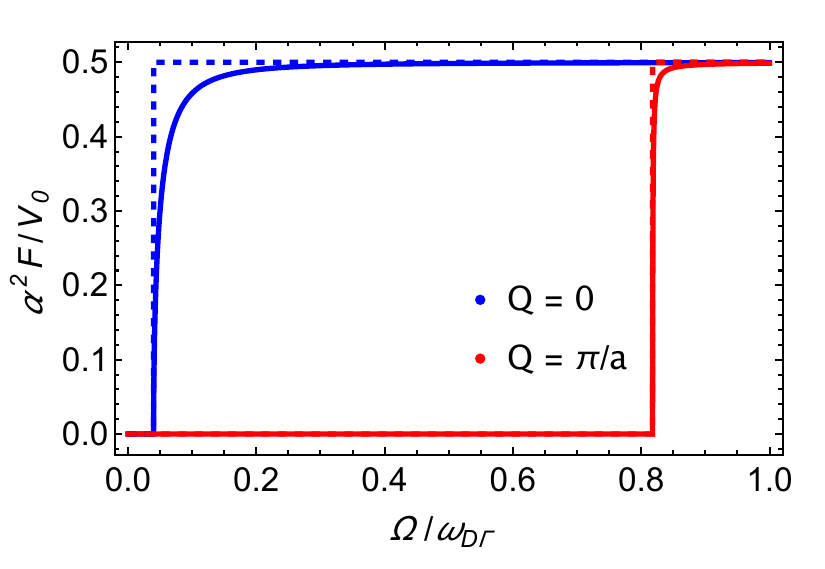}
    \caption{Plots comparison of the exact (solid lines) and approximated (dashed lines) Eliashberg function $\alpha^2F(\Omega)/V_0$ for ${ Q}=0$ (blue lines) and ${ Q}=\pi/a$ (red lines) in the soft phonon case ($\omega_{\Gamma,\text{TO} }=1$meV).}
   \label{fig:comparison}
\end{figure}

The expression in Eq.~\eqref{eq:alfaexact} is general and applies to any system whose electron-electron interaction is mediated by two transverse optical phonons with the dispersion of Eq.~\eqref{PhononDispersion}. For  $\omega_{\Gamma,\text{TO} }=0$ it is constant between a lower and an upper limit given by the $\Theta$ functions. 
For $\omega_{\Gamma,\text{TO} }>0$, the square root yields a rounded step as shown by the full lines in Fig.~\ref{fig:comparison}. Since the effective interaction Eq.~\eqref{Veffica} depends on an integral over frequency, for small $\omega_{\Gamma,\text{TO} }$, a good approximation is to substitute the rounded step by a sharp step,
\begin{equation}\label{elfunap}
    \alpha^2 F(\bm{ Q},\Omega) = \frac{V_0}{2}\Theta\big(\Omega-2\omega_{{\bm{ Q}}/{2}}\big)\Theta\big(2\omega_D({\bm{ Q}})-\Omega\big),
\end{equation}
as shown in Fig.~\ref{fig:comparison} by the dashed lines. 
Substituting this expression in Eq.~\eqref{Veffica} we obtain the effective two-phonon interaction,    \begin{equation}\label{poteffan}
     {V_\text{eff}\big(\bm{ Q}, i\Omega_n \big)} = \frac{V_0}2 \log\Bigg[\frac{\Omega_n^2+4\omega_{{\bm{ Q}}/{2}}^2}{\Omega_n^2+4\omega_D^2(\bm{ Q})}\Bigg].
\end{equation}
Performing the analytic continuation, we obtain the following effective interaction in the real frequency axis, 
\begin{equation}
     \begin{split}    
     {V_\text{eff}\big(\bm{ Q}, \omega\big)} & =
     V_0\Biggl\{-i\pi\mathop{\mathrm{sgn}}(\omega)\Theta\Bigg(\frac{\omega^2-4\omega_{{\bm{ Q}}/{2}}^2}{4\omega_D^2(\bm{ Q})-\omega^2}\Bigg)\\
     &+\log\abs{\frac{\omega^2-4\omega_{{\bm{ Q}}/{2}}^2}{\omega^2-4\omega_D^2(\bm{ Q})}}\Biggl\}.
     \end{split}
\end{equation}

\begin{figure}[tb]
 \centering  \includegraphics[width=1\linewidth]{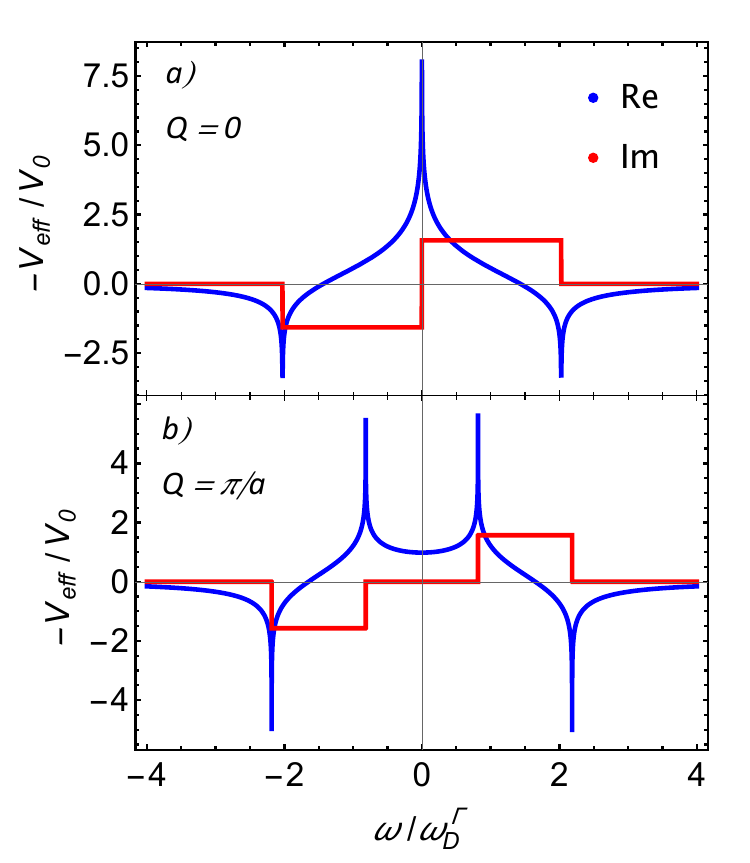}
    \caption{Plots of the real (blue line) and imaginary part (red line) of $-V_\text{eff}(\omega)/V_0$ in the acoustic-dispersion regime ($\omega_{\Gamma,\text{TO} }=0$ meV) for ${ Q}=0$ (a) and ${ Q}=\pi/a$ (b).}
    \label{fig:Veff1}
\end{figure}

We can expect that the maximum attraction occurs at the ferroelectric QCP, $\omega_{\Gamma,\text{TO} }\rightarrow0$. Indeed, in this case, the real part of the effective interaction has a singularity at zero frequency and zero momentum as shown by the blue line in  Fig.~\ref{fig:Veff1}(a). However, the singularity is only logarithmic, which, as can be anticipated, will not play an important role when integrated in frequency and momentum.  
For finite momentum, the singularity moves to finite frequency as shown in  Fig.~\ref{fig:Veff1}(b). The imaginary part has steps and is nonzero in the regions where the phonon DOS is finite. Each time the imaginary part has a discontinuity, the real part has a logarithmic divergence.

\subsection{Linearized Eliashberg equation and $T_c$}\label{DynLinEq}
To estimate $T_c$, we consider a single band, which, for simplicity, is taken as isotropic with a parabolic dispersion, 
\begin{equation}
    \xi_{{\bm{ k}}}=\xi_{{{ k}}}=\frac{{k}^2}{2m}-\mu,
\end{equation}
where $m$ is the effective electron mass.

Due to the sphericity of the Fermi surface and the isotropy of the effective interaction $V_\text{eff}\big(\bm{ Q}, i\Omega_n \big)$,  the renormalization functions and the gap function (defined in Eq.~\eqref{self-energy}) do not depend on the direction of $\bm{ k}$ but on its modulus only (i.e., $\Delta( \bm{k},i\omega_n)=\Delta( k,i\omega_n)$, etc.).
For simplicity, we neglect the frequency dependence of $Z({{ k}},i\omega_n)$ and $\chi({{ k}},i\omega_n)$ and evaluate them in the static limit. Furthermore, we neglect the imaginary part of the normal state electron self-energy. 
With these approximations, the two functions can be incorporated in the definition of $\xi_{\bm{ k}}$. 
Therefore, in the following, we take
\begin{equation}\label{neglfreqdep}
    Z({{ k}},i\omega_n)=1, \quad \chi({{ k}},i\omega_n)=0.
\end{equation}

The Eliashberg equations reduce to Eq.~\eqref{eleq2} and $T_c$ is determined by the linearized form,
\begin{equation}\label{LinearizedEq}
\begin{split}
    &\Delta( k,i\omega_n)=-\frac{T_c}{N}\\
    &\times\sum_{\bm{ k}'i\omega_{n'}}\frac{{V_\text{eff}(\abs{\bm{ k}-\bm{ k}'}, i\omega_n-i\omega_{n'})}}{\omega_{n'}^2+\xi_ {k'}^2}\Delta( k',i\omega_{n'}).
    \end{split}
\end{equation}

Converting the sums to integrals and setting 
$\abs{\bm{ k}-\bm{ k}'}^2={{ k}}^2+{{ k}'}^2-2{{ k}}{{ k}'}\cos\theta$, the angular integral can be done, allowing us to define the following dimensionless interaction,
\begin{equation}
    {\mathit{v}}({{ k}},{{ k}'},i\Omega_n) \equiv\frac{2}{V_0}\int_{-1}^{1} d(\cos\theta){V_\text{eff}(\abs{\bm{ k}-\bm{ k}'}, i\Omega_n)}.\label{eq:v}
\end{equation}

\begin{figure}[tb]
 \centering
\includegraphics[width=1\linewidth]{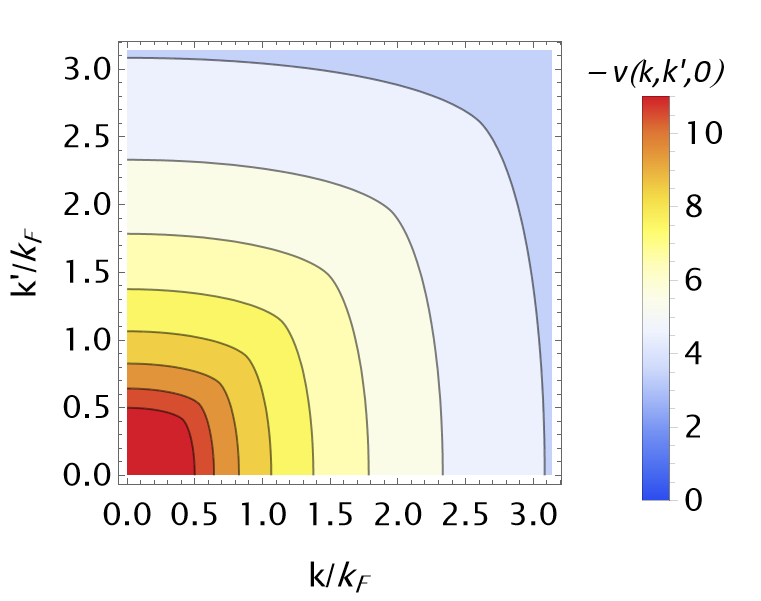}
    \caption{Contour plots of $-{\mathit{v}}({{ k}},{{ k}'},i\Omega_n=0)$ as a function of $k$, $k'$ in the acoustic-dispersion regime ($\omega_{\Gamma,\text{TO} }=0$) with $\omega_{D\Gamma}=50$ meV. 
    }
    \label{fig:Vmed}
\end{figure}

Performing the integrals, we obtain, 
\begin{equation}
\begin{split}\label{eq:v2}  
    &{\mathit{v}}({{ k}},{{ k}'},i\Omega_n)\\
    &=-\frac{c^2(k^2+k'^2)+4{\omega_{D \Gamma}^2}+{\Omega_n^2}}{2c^2kk'}\log\Bigg(\frac{A_{D\Gamma}^+({{ k}},{{ k}'},i\Omega_n)}{A_{D\Gamma}^-({{ k}},{{ k}'},i\Omega_n)}\Bigg)\\
    &+\frac{c^2(k^2+k'^2)+4{\omega_{\text{TO}\Gamma}^2}+{\Omega_n^2}}{2c^2kk'}\log\Bigg(\frac{A_{\text{TO}\Gamma}^+({{ k}},{{ k}'},i\Omega_n)}{A_{\text{TO}\Gamma}^-({{ k}},{{ k}'},i\Omega_n)}\Bigg)\\
    &-\log\Bigg(\frac{A_{D\Gamma}^+({{ k}},{{ k}'},i\Omega_n)}{A_{\text{TO}\Gamma}^+({{ k}},{{ k}'},i\Omega_n)}\frac{A_{D\Gamma}^-({{ k}},{{ k}'},i\Omega_n)}{A_{\text{TO}\Gamma}^-({{ k}},{{ k}'},i\Omega_n)}\Bigg),\\
    \end{split} 
\end{equation}
where
\begin{equation}
   \begin{split}
   &A_{D\Gamma}^\pm({{ k}},{{ k}'},i\Omega_n)= {c^2(k\pm k')^2+4{\omega_{D \Gamma}^2}+{\Omega_n^2}},\\
   &A_{\text{TO}\Gamma}^\pm({{ k}},{{ k}'},i\Omega_n)={c^2(k\pm k')^2+4{\omega_{\text{TO}\Gamma}^2}+{\Omega_n^2}}.
   \end{split}
\end{equation}

Figure~\ref{fig:Vmed} shows the angular-averaged static interaction as a function of the magnitude of the momentum of the two-scattering electrons for the system at the ferroelectric QCP ($\omega_{\Gamma,\text{TO} }=0$).
The attraction decreases smoothly as the momentum increases and has again a singularity at small momentum, behaving as,  
\begin{equation}
    \lim_{k,k'\rightarrow0}\left.{\mathit{v}}({{ k}},{k'},i\Omega_n)\right\rvert_{\omega_{\Gamma,\text{TO} }=0}=2\log\Bigg[\frac{
    \Omega_n^2}{
\Omega_n^2+4\omega_{D\Gamma}^2}\Bigg].
\end{equation}

It is convenient to define the following integral, 
\begin{equation}\label{dom}
    X(i\omega_n)
    =\int d{{ k}}{{ k}}^2\frac{{\Delta}({{ k}},i\omega_n)}{\omega_{n}^2+\xi_{{ k}}^2},
\end{equation}
where the limit of integration can be extended to infinity, as the gap function should decay at large momentum and the denominator behaves as $1/k^4$.

The linearized equation becomes,
\begin{equation}\label{eqlincondeltatilde}
\begin{split}  
    &X(i\omega_n)=-\frac{T_cV_0}{2(2\pi)^2}\\
    &\times\sum_{i\omega_{n'}}\int d{{ k}'}d{{ k}}\frac{ {({ k}}{{ k}')^2} {\mathit{v}}({{ k}},{{ k}'},i\omega_n-i\omega_{n'})}{{({\omega_{n'}^2+\xi_ {k'}^2})({\omega_{n}^2+\xi_ k^2})}}\Delta({{ k}'},i\omega_{n'}).\\
    \end{split}
\end{equation} 
 Assuming the solution is dominated by small Matsubara frequencies, we observe that  
for small $\omega_n$, $\omega_{n'}$, 
the term multiplying ${\Delta}({{ k}'},i\omega_{n'})$ in Eq.~\eqref{eqlincondeltatilde} is peaked at the Fermi wavevector (${{ k}}={{ k}'}=k_F$).  
Since $\mathit{v}({{ k}},{{ k}'},i\Omega_n)$ is smooth and its log-divergence at $k=k'=0$ for $\Omega_n=0$ is canceled by the factor $(kk')^2$, it is possible to make the following approximation
\begin{equation}\label{vfun}
    {\mathit{v}}({{ k}},{{ k}'},i\Omega_n)={\mathit{v}}(k_F,k_F,i\Omega_n).
\end{equation}
The assumption will be verified 
{\em a posteriori} showing that indeed $\Delta({{k}},i\omega_{n})$ is larger at small Matsubara frequencies. With this approximation, the interaction has become separable, and the equation can be solved. Notice that we are not assuming the pairing is finite only close to the Fermi wave-vector (as it is usually done when the Migdal theorem applies).  We are only evaluating the interaction matrix element at the dominant wave-vector and allowing for the pairing of all electrons in the Fermi sphere. 

Performing the momentum integral, we finally obtain the following linear eigenvalue problem, 
\begin{equation}\label{MatrixEquation}    
X(i{\omega}_n)=\sum_{i{\omega}_{n'}}M(i{\omega}_n,i{\omega}_{n'})X(i{\omega}_{n'}),
\end{equation}
where we have introduced the matrix element,
\begin{equation}\label{matrixelement}    
    M(i\omega_n,i\omega_{n'})= h(i\omega_n){\mathit{v}}(k_F,k_F,i\omega_n-i\omega_{n'}),
\end{equation}
given in terms of the dimensionless function
\begin{equation}\label{ffunction}
    \begin{split}
    h(i\omega_n)=&-
    \frac{i\lambda}{16}
    \frac{ \pi T_c}{\omega_n}\\
    &\times\bigg({\sqrt{-1-\frac{i\omega_n}{\epsilon_F}}}-{\sqrt{-1+\frac{i\omega_n}{\epsilon_F}}}\bigg),\\
    \end{split}
\end{equation}
with 
\begin{equation}\label{eq:lambda}
    \lambda
    =\frac{V_0}{2\pi^2W} \sqrt{\frac{\epsilon_F}W}=N_0V_0.
\end{equation}
Here $N_0$ is the density of states per spin at the Fermi level,
\begin{equation}\label{DOSElec}
    N_0\equiv\frac{\sqrt{\epsilon_F}}{2\pi^2W^{\frac{3}{2}}},
\end{equation}
with 
\begin{equation}\label{W}
    W=\frac{\hbar^2}{2ma^2}.
\end{equation}
Notice that $\lambda$ has the conventional BCS form. Similar to Eq.~\eqref{eq:v2}, it is convenient to 
work with dimensionless Matsubara frequencies by measuring all energies in units of $\epsilon_F$. Then one obtains two more dimensionless parameters, $\omega_{D\Gamma}/\epsilon_F$ controlling the degree of adiabaticity and $\omega_{\Gamma,\text{TO} }/\epsilon_F$ controlling the distance to the QCP.

To obtain the critical temperature, one has to solve the eigenvalue problem 
\begin{equation}\label{eigpro}
    \bar{\bar{M}} \bar{ X}=\alpha(T)\bar{ X}. 
\end{equation}
where  $\bar{\bar{M}}$ and $ \bar{ X}$ are a matrix and a vector in Matsubara space, respectively. 
As usual, $T_c$ is defined as the highest temperature for which $\alpha(T=T_c)=1$. 

Notice that since the matrix Eq.~\eqref{matrixelement} is given by the product of a real diagonal matrix and a real symmetric matrix, it can be diagonalized, and its eigenvalues are real.

Using the symmetry properties of the matrix $M$, it is possible to show that  $X(i\omega_n)= X(i\omega_{-n-1})$ and consider only non-negative Matsubara frequencies. In the computations below, we truncated the matrix to $0\le n,n'< N$, leading to an  $N\times N$ matrix with $N=200$. We checked that for this $N$ the computation converges to the asymptotic value.

Given the solution $X(i\omega_n)$,  the gap function can be obtained as
\begin{equation}\label{FromDtoDelta}    
    \Delta(k,i{\omega}_n)=-\frac{T_cV_0}{8\pi^2}\sum_{i{\omega}_{n'}}\mathit{v}({{ k}},{k_F},i\Omega_n)X(i{\omega}_{n'}),
\end{equation}
where we restore the full momentum dependence of the interaction on the first argument. This ensures that the gap function vanishes when the pairing vanishes at large momentum.

Again, we emphasize that the resulting gap function is very different from the conventional Migdal-Eliashberg theory, where the gap function is different from zero only for quasi-particles near the Fermi surface. Here, because we allow for high-frequency phonon excitations, all electrons can potentially participate in the condensate.

\section{Numerical Results}
\label{TheoryControl}
Since superconductivity in STO occurs at low temperatures, we approximated the chemical potential with the zero temperature Fermi energy, $\mu=\epsilon_F$.

\begin{figure}
    \centering
\includegraphics[width=\linewidth]{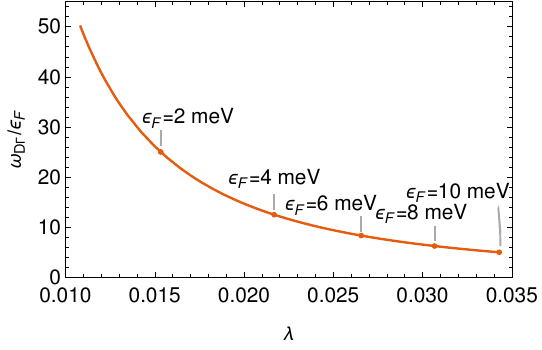}
  \caption{Parametric plot of the ratio $\omega_{D\Gamma}/\epsilon_F$ vs. $\lambda$ fixing parameters as in Table~\ref{tab:par} and varying the Fermi energy in the range $1-10$~meV as relevant for STO.}
    \label{fig:wddl}
\end{figure}

\begin{table}[h]
\caption{Reference values of key parameters for two-phonon pairing in STO. The Slater reduced mass can be computed in terms of the O and Ti atomic masses as $\mu_S^{-1}=(3M_O)^{-1}+M_{Ti}^{-1}$.\label{tab:par}}
\begin{center}
\begin{ruledtabular}
\begin{tabular}{c|c c c}
Quantity &  Value & Source  \\
\hline
$m$ & $2m_e$ & Refs.~\cite{Lin2014,McCalla2019N}\\
W &    123~\text{meV}. & Eq.~\eqref{W}\\
$\mu_S$  & $24$~u &  \\
$\hbar c/a$ (TO1)     & 13 meV & Ref.~\cite{Shirane1969N}\\ 
$\omega_{D\Gamma}$ & $50$~meV &Eq.~\eqref{OmegaD0} \\
$\alpha$ & $40~{\text{eV/\AA}}^2$ & Table~\ref{tab:alpha} \\
$V_0 $ & 290~meV  \footnote{\label{not:f2}Upper bound corresponding to  $\bar f=2$. }                   &Eq.~\eqref{V0}  \\
$\Lambda$  &0.08\footref{not:f2} & Eq.~\eqref{eq:Lambda}       
\end{tabular}
\end{ruledtabular}
\end{center}

\label{tabellaconfronti}
\end{table}

Table~\ref{tab:par} summarizes key parameters for two-phonon pairing in STO. To evaluate $W$, we took the electron effective mass $m$ equal to twice the mass of the electron at rest $m_e$, chosen to match the values from quantum oscillation experiments 
\cite{Lin2014} and specific heat~\cite{McCalla2019N}. The velocity characterizing the two TO modes is taken from Ref.~\cite{Shirane1969N}. 
In the one-band regime, the doping level is given by
\begin{equation}
    x=\frac{k_F^3a^3}{3\pi^2}=\frac1{3\pi^2}\left(\frac{\epsilon_F}W \right)^{3/2}.
\end{equation}

It is instructive to see what the parameters of the model are for typical densities in STO. 
Both $\lambda$ [Eq.~\eqref{eq:lambda}] and the adiabatic parameter,  $\omega_{D\Gamma}/\epsilon_F$ depend on the Fermi energy. Figure~\ref{fig:wddl} shows their parametric plot, as the Fermi energy is changed in a 
relevant range for superconductivity in STO, 1~meV~$< \epsilon_F < 10$~meV~\cite{Lin2014}.

Despite the very large value of $\alpha$ (the two-phonon deformation potential), the dimensionless electron-phonon coupling is small. This can be traced back to the fact that the cutoff frequency appears in the denominator of Eq.~\eqref{V0} in contrast to the soft phonon frequency in the toy model Eq.~\eqref{effpotquad}. Therefore, the effective interaction gets decreased by a factor $(1/50)^3\sim 10^{-5}$ with respect to the naive estimate.

Since we are interested in the behavior of $T_c$ for a range of doping, it is useful to define a second coupling constant that characterizes the material independently of the doping level [c.f. Eq.~\eqref{eq:lambda}], 
\begin{equation}\label{eq:Lambda}
    \Lambda\equiv \lambda\sqrt{\frac{\omega_{D\Gamma}}{\epsilon_F}}= \frac{{V}_0}{{2\pi^2}W}\sqrt{\frac{{\omega}_{D\Gamma}}{W}}.
\end{equation}
This yields the value shown in Table~\ref{tab:par}, which again indicates that STO is in the weak coupling regime. 

In the following, we assume the system to be at the QCP independently of the value of doping, i.e.,  we consider the acoustic-dispersion regime ($\omega_{\Gamma,\text{TO} }=0$) as this yields the highest possible critical temperature. 

\begin{figure}[tb]
 \centering
\includegraphics[width=1.0\linewidth]{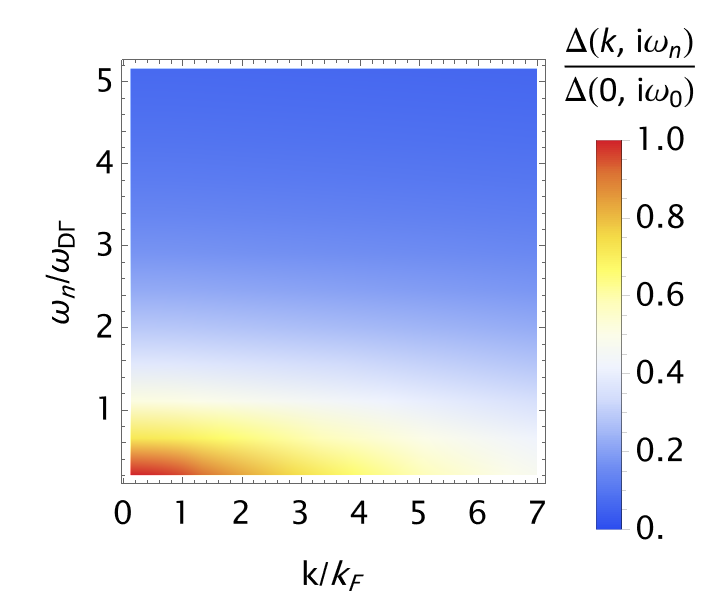}
    \caption{False color plot of $\Delta(k,i\omega_n)/\Delta(0,i\omega_0)$ at the QCP ($\omega_{\Gamma,\text{TO} }=0$) for $\lambda=0.41$ and $\omega_{D\Gamma}/\epsilon_F$=2. 
}
    \label{fig:Dk}
\end{figure}

Figure~\ref{fig:Dk} shows an example of the momentum and Matsubara frequency dependence of the gap function,  $\Delta(k,i\omega_n)$, for a moderate value of coupling constant $\lambda=0.4$.  As anticipated previously, we find that the gap function is larger at $\omega_{n=0}$ and decays rapidly as the Matsubara frequency increases for every value of $k$. This further justifies the approximation made in Eq.~\eqref{vfun}, which was based on the dominance of the low Matsubara behavior. It is also possible to see that the gap function has its largest value at $k=0$ and decreases monotonously with momentum.


\begin{figure}[tb]
\includegraphics[width=1.0\linewidth]{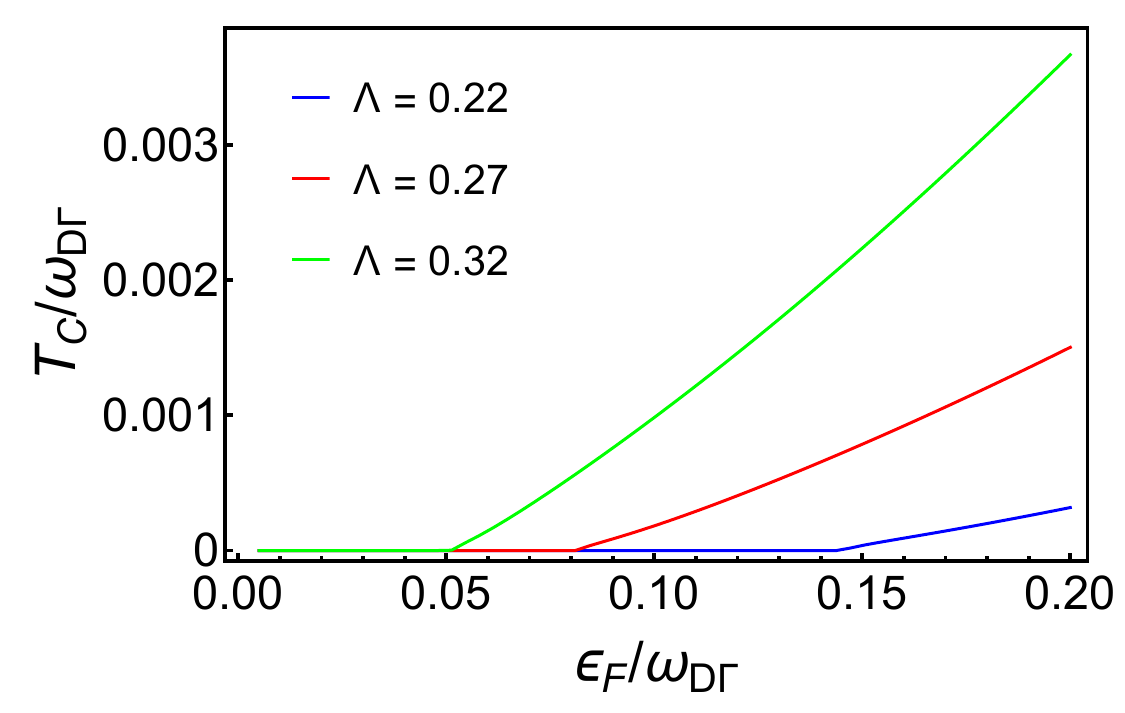}    \caption{${T}_c/\omega_{D\Gamma}$ vs $x$ in the acoustic case ($\omega_{\Gamma,\text{TO} }=0$) for $\Lambda=0.22$ (blue line), $\Lambda=0.27$ (red line), $\Lambda=0.32$ (green line). ${\omega}_{D\Gamma}/W=0.4$. ${T}_c$ has been plotted using $60$ different values for ${\epsilon_F}/{\omega_{D\Gamma}}$. 
    }
    \label{fig:TcVsx}
\end{figure}

In Fig.~\ref{fig:TcVsx} 
we report $T_c/\omega_{D\Gamma}$ as a function of the electron doping $x$ for different values of $\Lambda$. We have fixed the ratio $\omega_{D\Gamma}/W=0.4$, which is the value for STO, but we have to take values of $\Lambda$ larger than the one in Table~\ref{tab:par}, to get $T_c$ of order $0.6$~K corresponding to $T_c/\omega_{D\Gamma}\approx0.001$.  $T_c$ increases monotonically with $x$ and $\Lambda$.  
The plot shows that to have a sizable $T_c$ for typical dopings in STO, 
$\Lambda$ should be roughly a factor of three larger than the value, in Table~\ref{tab:par}. 

\begin{figure}
    \centering
\includegraphics[width=\linewidth]{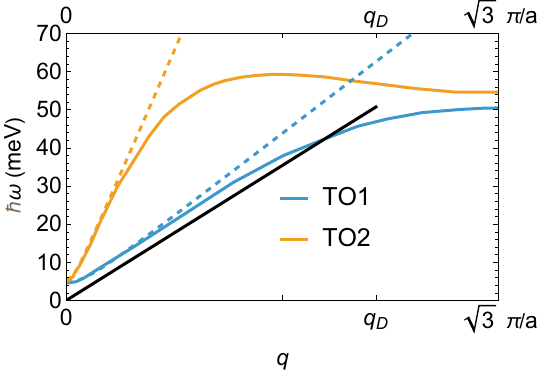}
  \caption{The two lowest TO modes according to the computations of Ref.~\cite{Verdi2023}. Data points were taken along the $\Gamma$-R direction of their Fig.~2, ignoring obvious avoided crossings due to hybridization with other modes. The dashed line are fits with Eq.~\eqref{PhononDispersion} yielding $\omega_{\text{TO}\Gamma}=4.7 (5)$~meV and $\hbar c/a=16 (48)$meV for the TO1 (TO2) mode.  For comparison, the black full line shows the dispersion relation used in our computations of $T_c$  plotted until $q_D$, the isotropic cutoff in momentum space.}
    \label{fig:wdq}
\end{figure}

Given the strong sensitivity of $V_0$ to the cutoff frequency, we can examine if details of the model can allow the needed boost. In the present computation, we have taken a linear phonon dispersion until the energy cutoff [Eq.~\eqref{PhononDispersion}], corresponding roughly to the zone boundary. 
In more recent computations of phonons~\cite{Verdi2023}, which are in good agreement with experiments, the two lowest TO branches bend noticeably and saturate at 50 meV near the zone boundary, as shown in Fig.~\ref{fig:wdq}. Notice that the more accurate dispersion in almost all of the Brillouin zone comprises modes of higher energy than our model, which therefore would lead to an even smaller $T_c$. 

More importantly, our computation shows that two-phonon superconductivity is not dominated by the physics of the soft mode but by the details of the high-energy part of the spectrum. This is a consequence of the fact that the Eliashberg function Eq.~\eqref{eliashbergfnc} is an average of the two-phonon processes over all the Brillouin zone.

\section{Summary and conclusions}

Our computations show that the values of electron-phonon interaction relevant for the two-phonon mechanism proposed by Ngai are too small to explain superconductivity in STO. To be on the safe side, we have taken values of the two-phonon deformation potential $\alpha$ larger than the experimental estimates, and a phonon frequency smaller than the more accurate computations we know;  still, $T_c$ appears negligibly small. Keeping our dispersion relation, a 30\% reduction of the cutoff frequency translates into a factor of 3 increase of $\Lambda$, which then would be in the correct range.

This, however, stresses another fact that makes the two-phonon mechanism unlikely to explain the physics of pairing in STO. 
The main motivation to explore the Ngai mechanism of superconductivity is that experimentally, the phonon softening appears to be closely related to superconductivity. On the other hand, 
our computations show that the two-phonon mechanism is not dominated by the physics of the soft mode but by the details of the high-energy part of the spectra. This is a consequence of the fact that the Eliashberg function Eq.~\eqref{eliashbergfnc} is an average of the two-phonon processes over all the Brillouin zone. 
Indeed, we found that the enhancement of the effective interaction due to the softening is modest, as it goes as the logarithm of the ratio of the Debye frequency to the soft phonon dispersion, as seen in Eq.~\eqref{poteffan}.  It is still likely
that the two-phonon mechanism contributes to $T_c$, as recently proposed for KTO~\cite{Norman2026} and STO~\cite{Saha2025}. Clearly, the same can be said of all phonons contributing significantly to the electron-phonon coupling~\cite{Esswein2023N}.  

Still, the arguments of Sec.~\ref{toymod} suggest that the two-phonon mechanism is potentially very promising and can even lead to high-$T_c$ superconductivity. Therefore, an interesting question is what characteristic a material should have to profit from this mechanism. Clearly, the more important property to make the mechanism successful is the soft mode behavior to persist in all the Brillouin zone instead of being confined to a small region. In this case, the arguments of Sec.~\ref{toymod} would apply, leading to a strong attraction. 

A potentially interesting system is hafnia (\ch{HfO2}). This system has many competing structural phases and a very unusual ferroelectricity that survives in the ultrathin limit. Among the different scenarios that have been proposed to explain scale-free ferroelectricity in hafnia, one includes the presence of flat polar-phonon bands~\cite{Lee2020,Qi2025}. Superconductivity in doped hafnia has been considered theoretically with a conventional mechanism~\cite{Duan2023}. Extending this study to a two-phonon channel can be very interesting. 

Our computations also suggest examining skutterudites in which some atoms are in cages larger than their ionic radii, producing Einstein-like soft phonons~\cite{Hermann2003}. Some are superconducting~\cite{Maple2008,Mizukami2020N,Sundaramoorthy2023N}, which makes this another interesting family for testing the Ngai mechanism and searching for optimal two-phonon superconductivity. 



\section*{Acknowledgements}
We are in debt to Dirk van der Marel for stimulating discussions.
We acknowledge the CINECA award under the ISCRA initiative Grants No. HP10CPHFAR, No. HP10CMZFMM and No. HP10CM2RK0, for the availability of high-performance computing resources and support.
M.N.G. is supported by the grants RYC2021-031639-I and PID2023-153277NB-I00 funded by MCIN/AEI/10.13039/501100011033. J.L acknowledges support from the Italian MUR through PRIN project  2022WS9MS4 "COINEX". 


\appendix

\section{Relation between tetragonal unit cell and pseudocubic unit cell}
\begin{figure}
    \centering
    \includegraphics[width=\linewidth]{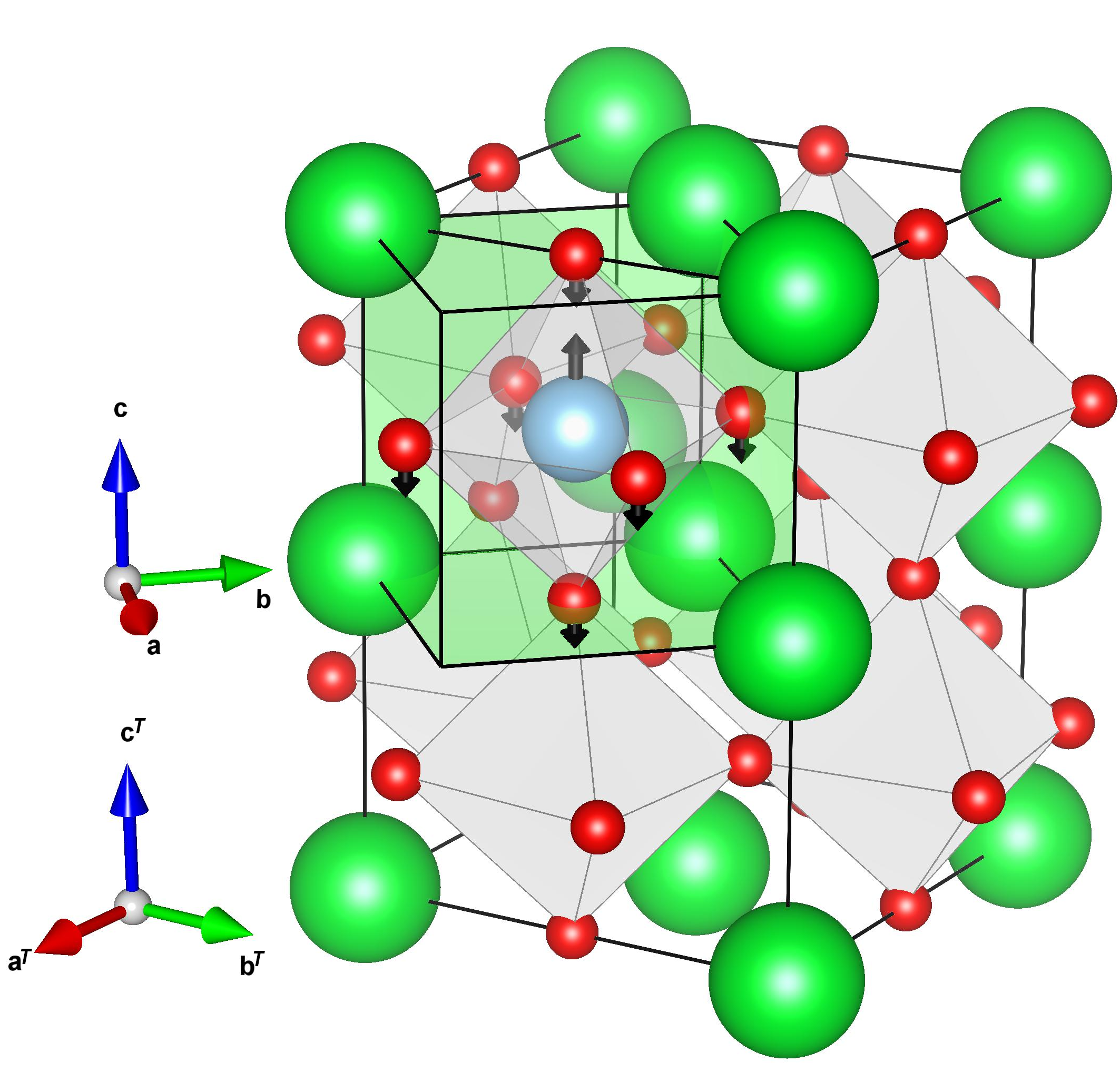}
    \caption{Schematic of the relation between the tetragonal unit cell and the pseudocubic unit cell (green shade). We show the lattice displacements of the Slater mode only inside the pseudocubic unit cell, but they repeat periodically in the whole system. The left-hand insets show the orientation of the axes for the cubic cell (top) and the tetragonal unit cell (bottom). }
    \label{fig:tetra}
\end{figure}

Figure~\ref{fig:tetra} shows the relation between the pseudocubic unit cell used to index directions in the main text and the tetragonal unit cell, which corresponds to the crystal structure below 105~K.

\label{app}
\clearpage




\bibliography{MasterBib,library_lorenzana}






\end{document}